\documentclass{aa}  

\usepackage{graphicx}
\usepackage{natbib}
\bibpunct{(}{)}{;}{a}{}{,} 

\usepackage{txfonts}

\usepackage{xcolor,colortbl}
\definecolor{Gray}{gray}{0.85}

\begin{document}

	\title{The Tucana dwarf spheroidal galaxy: not such a massive failure after all \thanks{Based on observations made with ESO telescopes at the La Silla Paranal Observatory as part of the programs 091.B-0251, 69.B-0305(B) and 095.B-0133(A).
		}
	}
	
	\subtitle{}
	
	\author{S. Taibi \inst{1}\fnmsep \inst{2} \fnmsep \thanks{\email{staibi@iac.es}}, 
		G. Battaglia\inst{1}\fnmsep \inst{2}, 
		M. Rejkuba\inst{3},
		R. Leaman\inst{4},
		N. Kacharov\inst{4},
		G. Iorio\inst{5},
		P. Jablonka\inst{6,7},
		M. Zoccali\inst{8,9}
	}
	
	\institute{Instituto de Astrofisica de Canarias, C/ Via Lactea s/n, E-38205, La Laguna, Tenerife, Spain
		\and
		Departamento de Astrofisica, Universidad de La Laguna, E-38205, La Laguna, Tenerife, Spain
		\and
		European Southern Observatory, Karl-Schwarzschild Strasse 2, D-85748 Garching, Germany
		\and
		Max Planck Institute for Astronomy, K\"{o}nigstuhl 17, D-69117 Heidelberg, Germany
		\and
		Dipartimento di Fisica e Astronomia ``G. Galilei'', Universit\`{a} di Padova, vicolo dell’Osservatorio 3, 35122 PD, Italy 
		\and
		Institute of Physics, Laboratory of Astrophysics, Ecole Polytechnique Federale de Lausanne (EPFL), 1290 Sauverny, Switzerland
		\and
		GEPI, CNRS UMR 8111, Observatoire de Paris, PSL Research University, F-92125, Meudon, Cedex, France
		\and
		Instituto de Astrofisica, Pontificia Universidad Catolica de Chile, Av. Vicu\~{n}a Mackenna 4860, 782-0436 Macul, Santiago, Chile
		\and
		Millennium Institute of Astrophysics, Av. Vicu\~{n}a Mackenna 4860, 782-0436 Macul, Santiago, Chile
		}
	
	\date{Received M DD, YYYY; accepted M DD, YYYY}
	
	
	\abstract
	{Isolated Local Group (LG) dwarf galaxies have evolved most or all of their life unaffected by interactions with the large LG spirals and therefore offer the opportunity to learn about the intrinsic characteristics of this class of objects.}
	{Here we explore the internal kinematic and metallicity properties of one of the three isolated LG early-type dwarf galaxies, the Tucana dwarf spheroidal. This is an intriguing system, as it has been found in the literature to have an internal rotation of up to 16 km\,s$^{-1}$, a much higher velocity dispersion than dwarf spheroidals of similar luminosity, and a possible exception to the too-big-too-fail problem.}
	{We present the results of a new spectroscopic dataset from the Very Large Telescope (VLT) taken with the FORS2 instrument in the region of the Ca~II triplet for 50 candidate red giant branch stars in the direction of the Tucana dwarf spheroidal.
	This yielded line-of-sight (l.o.s.) velocity and metallicity ([Fe/H]) measurements of 39 effective members, which doubles the number of Tucana's stars with such measurements. In addition, we re-reduce and include in our analysis the other two spectroscopic datasets presented in the literature, the VLT/FORS2 sample by \citet{Fraternali2009} and the VLT/FLAMES one by \citet{Gregory2019}.}
	{We measure a l.o.s. systemic velocity of $180\pm1.3$ km\,s$^{-1}$, consistently across the various datasets analyzed, and find that a dispersion-only model is moderately favored over models accounting also for internal rotation. Our best estimate of the internal l.o.s. velocity dispersion is 6.2$_{-1.3}^{+1.6}$ km\,s$^{-1}$, much smaller than the values reported in the literature and in line with similarly luminous dwarf spheroidals; this is consistent with NFW halos of circular velocities $<30$ km\,s$^{-1}$. Therefore, Tucana does not appear to be an exception to the too-big-to-fail problem nor to live in a dark matter halo much more massive than those of its sibling.
	As for the metallicity properties, we do not find anything unusual; there are hints of the presence of a metallicity gradient but more data are needed to pin its presence down.
	}
	{}
	
	
	\titlerunning{Chemo-kinematics of the Tucana dSph}
	\authorrunning{S. Taibi et al.}
	
	\maketitle
	%
	
	\section{Introduction} \label{sec:intro}
	
	Dwarf galaxies are the least massive and yet the most dark-matter dominated galactic systems observed \citep*[e.g.][]{Mateo1998,Battaglia2013,Walker2013}. In the Local Group (LG) the nearest ones to the largest spirals, i.e. the Milky Way (MW) and M31, are gas-poor dwarf spheroidal systems (dSphs) with no on-going star formation \citep*{Tolstoy2009}. Although they share the same morphology, their full star formation histories show complex evolutionary pathways \citep{Gallart2015}.
	
	Due to their small masses, the formation and evolution of these galaxies can potentially be strongly influenced by environmental effects. The orbital properties of the dwarf galaxies satellites of the MW, obtained by integrating back in time their present-day systemic motion \citep[see e.g.][for Gaia-DR2 based determinations]{Helmi2018,Fritz2018,Simon2018}, are consistent with repeated pericentric passages for several of these objects. In practice, (unknown) factors such as the triaxiality of the MW's potential or interactions between satellites, among others, introduce significant uncertainties in the reconstruction of their full orbital history \citep[e.g.][]{Lux2010}, but comparison with the properties of dark matter sub-halos around MW-sized hosts do suggest that most MW satellites fell in the MW halo at intermediate-to-early times \citep[see][]{Rocha2012,Wetzel2015}. A large body of works has therefore focused on exploring the impact of tidal and/or ram-pressure stripping caused by a MW-sized host onto various properties of the dSph satellites of the MW (\citealp[see e.g.][]{Piatek+Pryor1995,Mayer2001a,Mayer2001b,Mayer2006,Read2006,Munoz2008,Klimentowski2009,Kazantzidis2011,Pasetto2011,Battaglia2015,Iorio2019}). 
	However, as recently showed by \citet{Hausammann2019}, the ram-pressure stripping induced by a host halo has its limits in the actual quench and gas depletion of a dSph.
	Internal effects, like stellar feedback due to episodic star formation and supernova-driven winds, play an important role too \citep[see e.g.][]{Sawala2010,Bermejo-Climent2018,Revaz+Jablonka2018}. 
	Recent hydro-dynamic cosmological simulations have shown, indeed, that both internal and environmental mechanisms are necessary to reproduce the observed properties of the LG dwarf galaxies \citep[see e.g.][]{Brooks+Zolotov2014,Garrison-Kimmel2014,Wetzel2016,Sawala2016}. 
	Stellar feedback, for example, was also found to be an important ingredient in enhancing tidal-stirring effects \citep{Kazantzidis2017}. 
	
	Given the multitude of physical mechanisms affecting low mass galaxy evolution, data on the ages, chemical abundances, spatial distribution and kinematics of the stellar component of LG dwarf galaxies are needed to understand the observed diversity of these systems.  
	Detailed observations of MW satellites have been accumulated over the past years including large spectroscopic datasets \citep[e.g.][]{Tolstoy2004,Battaglia2006,Battaglia2008b,Battaglia2011,Walker2009b,Kirby2011,Lemasle2012,Lemasle2014,Hendricks2014,Spencer2017}. The observations showed little to no signs of internal rotation \citep[e.g.][]{Battaglia2008b,Walker2008,Leaman2013,Wheeler2017}, pronounced radial metallicity gradients \citep[e.g.][]{Battaglia2006,Battaglia2011, Walker2009b,Kirby2011,Leaman2013} and multiple populations with different chemo-dynamical properties \citep[e.g.][]{Tolstoy2004,Battaglia2006,Battaglia2008b}. Nonetheless, only few simulations have explored the link between the formation scenarios and the internal kinematic and chemical status of dwarf galaxies \citep[][]{Revaz2009,Schroyen2013,Revaz+Jablonka2018}.
	
	It is likely that the MW influenced at least in part the evolution of its satellites. Therefore, isolated dSphs represent a valuable tool to understand the intrinsic properties of the systems that have spent all or most of their life in a more benign environment. They are crucial for our understanding of the possible formation scenarios. In the LG, just a handful of dSphs are found in isolation: namely And~XVIII, Cetus and Tucana. Although it is probable, based on their line-of-sight (l.o.s.) systemic velocity, that in the past they may have interacted once with the MW or M31 (\citealp{Lewis2007,Fraternali2009}; but see also \citealp{Sales2010} and \citealp{Teyssier2012}), the fact that they spent most of their life in isolation makes them ideal targets to contrast with the satellite dwarfs. 
	
	This work is part of a larger body of studies aiming at improving our knowledge of the observed properties of a selected sample of isolated LG dwarf galaxies -- Phoenix, \citep{Kacharov2017}; Cetus, \citep{Taibi2018}; Aquarius, \citep{Hermosa-Munoz2019} -- mainly exploiting VLT/FORS2 multi-object spectroscopic data. 
	Here we focus on the Tucana dSph.
	
	Tucana is an early-type dwarf galaxy found in extreme isolation with a heliocentric distance of $D_\odot = 887\pm49$~kpc \citep{Bernard2009}, which places it at more than 1~Mpc away from M31 and the LG center, with only the Phoenix dwarf found within $\sim$500 kpc from it \citep{McConnachie2012}. Photometric observations have showed that the galaxy is mainly old and metal-poor, with an extended horizontal branch (HB) and a population of variable stars (see e.g. \citealp{Saviane1996,Bernard2009}). The structural analysis by \citet{Saviane1996} showed a highly flattened system ($e\sim0.5$) with a surface density profile well described by an exponential fit. 
	The recovery of the full star formation history (SFH) from deep HST/ACS observations reaching the oldest main sequence turn-off by \citet{Monelli2010}, showed that Tucana formed the majority of its stars more than 9~Gyr ago. It experienced a strong initial period of star formation (SF) starting very early on ($\sim$13~Gyr ago). 
	Tucana harbours at least two stellar sub-populations based on observed splitting of the HB, double red giant branch (RGB) bump, and the luminosity-period properties of the RR-Lyrae, which imply that this system experienced at least two early phases of SF in a short period of time. Using the same HST/ACS dataset \citet{Savino2019} refined the HB analysis showing that Tucana experienced two initial episodes of sustained SF followed by a third less intense, but more prolonged one, ending between 6 and 8 Gyr ago. The spatial analysis of the same dataset indicates the presence of a population age gradient inside $\sim 4\,R_e$ \citep{Monelli2010,Hidalgo2013,Savino2019}.
	
	The first spectroscopic study of individual stars in the Tucana dSph was conducted by \citet{Fraternali2009} with the VLT/FORS2 instrument obtaining a relatively small sample of $\sim$20 RGB probable member stars. They reported the systemic velocity and velocity dispersion values ($\bar{v}_{\rm sys}=194.0\pm4.3$ km\,s$^{-1}$ and $\sigma_{\rm v}=15.8_{-3.1}^{+4.1}$ km\,s$^{-1}$) for the galaxy, together with the presence of a maximum rotation signal of $\sim$16 km\,s$^{-1}$. They also determined a mean metallicity  value of [Fe/H] $=-1.95\pm0.15$~dex with a dispersion of $0.32\pm0.06$~dex.
	The determination of the systemic velocity ruled out an association with a nearby HI cloud, confirming that Tucana is devoid of neutral gas (down to a HI mass of $1.5\times10^4\,M_\odot$), and in addition is moving away from the LG barycenter \citep{Fraternali2009}. If bound, Tucana has not reached its apocenter yet. Indeed, it is possible that a past interaction between Tucana and the MW happened around 10~Gyr ago (roughly coinciding with the major drop in its SF; \citealp{Sales2007,Fraternali2009,Teyssier2012}). 
	
	Recently a new spectroscopic study of Tucana's RGB stars has been conducted by \citet{Gregory2019} using VLT/FLAMES data. Their sample of probable members is slightly larger than that of \citet{Fraternali2009}, but covers a much more extended spatial area (up to $\sim10\,R_{\rm e} \sim2\,R_{\rm tidal}$). Their velocity dispersion value ($\sigma_{\rm v}=14.4_{-2.3}^{+2.8}$ km\,s$^{-1}$) obtained for 36 probable members is similar to that of \citet{Fraternali2009}, although their systemic velocity shows a significant offset ($\bar{v}_{\rm sys}=216.7_{-2.8}^{+2.9}$ km\,s$^{-1}$); they also detect a velocity gradient of $k=7.6_{-4.3}^{+4.2}$ km\,s$^{-1}$\,arcmin$^{-1}$ along the optical major axis. Performing a dynamical modeling of Tucana's kinematic properties based on the FLAMES/GIRAFFE l.o.s. velocities of the probable members, the authors found a massive dark matter halo with a high central density. The implied dark matter halo mass profile is much denser than the other dwarfs of the LG, making Tucana the first exception of the \textit{too-big-to-fail} problem \citep[see e.g.][]{Boylan-Kolchin2012}. 
	In fact, the pure N-body simulations in the framework of $\Lambda$ cold dark matter ($\Lambda$-CDM) predict that the dwarf galaxies of the LG should live in denser halos than those inferred from observations. The fact that Tucana is found to reside in such massive halo in agreement with $\Lambda$-CDM predictions seems to indicate that during its evolution it has been able to maintain its initial content of dark matter, independently of the internal and environmental mechanisms that have driven its evolution. 

	Motivated by the unique internal kinematics of Tucana and the importance of isolated dwarfs in disentangling evolutionary processes, in this study we present results from a new investigation of the kinematic and chemical properties of the stellar component of the Tucana dSph. We have analyzed a new dataset of multi-object spectroscopic observations of 50 individual RGB stars taken with the VLT/FORS2 instrument targeting the near-IR wavelength region of the Ca~II triplet (CaT) lines. 
	To understand how systematics may influence the key results of our study, we have further re-reduced the original datasets presented in \citet{Fraternali2009} and \citet{Gregory2019} and performed a combined analysis together with our own data. In this work we present an in-depth and homogeneous analysis of all currently available spectroscopic data for the Tucana dSph.
	
	The	article is structured as follows. In Sect.~\ref{sec:acqred} we present the data acquisition and reduction processes for all the dataset analyzed in this work. Section~\ref{sec:measurements} is dedicated to the determination of the l.o.s. velocity and metallicity measurements. In Sect.~\ref{sec:membership} we describe the criteria applied to select likely member stars in the different dataset we present here. 
	Section~\ref{sec:kinematics} shows the results from the kinematic analysis, presenting the determination of the galaxy systemic velocity and velocity dispersion, alongside with the search of a possible rotation signal and the implication for the dark matter halo properties of Tucana. In Sect.~\ref{sec:chemical} we describe the determination of metallicities ([Fe/H]) and the subsequent chemical analysis. Finally, Sect.~\ref{sec:summary} is dedicated to the summary and conclusions, while in the appendixes we report the detailed comparison between the measurements obtained for our dataset with those reported in the literature, along with supplementary material from the kinematic analysis.
	
	\begin{table}
		\caption{Parameters adopted for the Tucana dSph: the coordinates of the galaxy's optical center; the ellipticity; the position angle; the core, tidal and half-light geometric radii; the stellar luminosity in \textit{V}-band; the tip of the red giant branch magnitude in \textit{I}-band; the average reddening; the heliocentric distance; the chemo-kinematic parameters obtained in this work, i.e. the systemic velocity, the velocity dispersion, the median metallicity and the intrinsic metallicity scatter.
		}             
		\label{table:1}      
		\centering          
		\begin{tabular}{l c c c}    
			\hline\hline
			Parameter & Units & Value  & Ref. \tablefootmark{$\star$}\\ 
			\hline           
			$\alpha_{\rm J2000}$ & & $22^h41^m49.6^s$ & (1) \\
			$\delta_{\rm J2000}$ & & $-64^{\circ}25'10''$ & (1) \\
			$\epsilon$\tablefootmark{a} & & 0.48$\pm$0.03 & (2) \\
			P.A. & deg & $97\pm2$ & (2) \\
			$R_{\rm core}$  & arcmin (pc) & $0.7\pm0.1$ ($955 \pm 139$) & (2) \\
			$R_{\rm tidal}$ & arcmin (pc) & $3.7\pm0.5$ ($181 \pm 28$)  & (2) \\
			$R_{\rm e}$     & arcmin (pc) & $0.8\pm0.1$ ($206 \pm 28$)  & (2) \\
			$L_{\rm V}$ & $10^5 M_\odot$ & $5.5\pm1.5$ & (2) \\
			$I_{\rm TRGB}$ & & $20.7\pm0.15$ & (2) \\
			E(B-V) & & 0.031 & (3) \\
			$D_\odot$ & kpc & $887\pm49$ & (3) \\
			$\bar{v}_{\rm sys}$ & km\,s$^{-1}$& $180\pm1.3$ & (4)\\
			$\sigma_{\rm v}$ & km\,s$^{-1}$& 6.2$_{-1.3}^{+1.6}$ & (4) \\
		    ${\rm [Fe/H]}$ & dex & $-1.58$ & (4)\\
		    $\sigma_{\rm [Fe/H]}$ & dex & 0.39 & (4)\\
			\hline        
		\end{tabular}
		\tablefoot{\tablefoottext{a}{$\epsilon=1-b/a$} 
			\tablefoottext{$\star$}{\textbf{References: }(1) \cite{Lavery+Mighell1992}; (2) \cite{Saviane1996}; (3) \cite{Bernard2009}; (4) this work.} 
		}
	\end{table}
	
	\section{Data acquisition and reduction processes} \label{sec:acqred}
	
	\subsection{The P91 FORS2 dataset} \label{subsec:data-fors2}
	
\begin{figure*}
		\centering
		\includegraphics[width=\hsize]{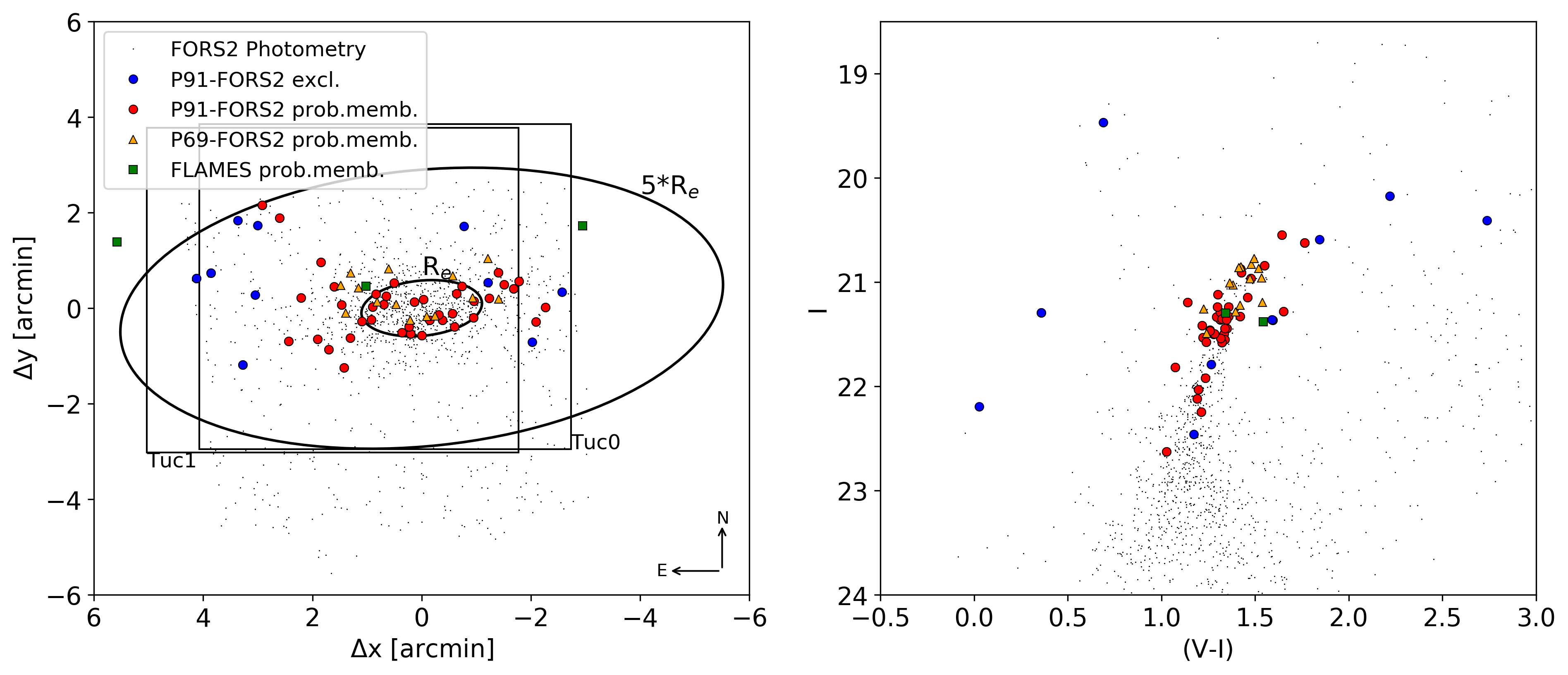}
		\caption{Spatial distribution (left) and color-magnitude diagram (right) of stars along the line-of-sight to the Tucana dSph. Black points represent the objects classified as stars in the FORS2 photometric data (see main text); red and blue dots indicate the P91-FORS2 MXU targets classified as probable members (i.e. with $P>0.05$) and non-members, respectively. Yellow triangles and green squares represent the probable member stars from the P69-FORS2 and FLAMES datasets, respectively, which were added to the P91-FORS2 and analyzed through the text.
			The 2 observed FORS2 pointings are represented as large squares, while the ellipses denote the galaxy half-light radius and the spatial extension of the dataset (i.e. up to $5\times R_e \sim R_{\rm tidal}$). We note that the photometric data are not corrected for reddening.
		}
		\label{Fig:Targ}
	\end{figure*}

	The primary dataset analyzed in this work was obtained with the FORS2 instrument mounted at the UT1 (Antu) of the Very Large Telescope (VLT) at the ESO Paranal observatory. Observations were taken in service mode over several nights between July 2013 and July 2014 as part of the ESO program 091.B-0251, PI: M. Zoccali (see Table~\ref{table:2}). The instrument was used in multi-object spectroscopic mode (MXU), which allows the observer to employ exchangeable masks with custom-cut slits. We used pre-imaging FORS2 photometry taken in Johnson V- and I-band, to allocate slits with stars having colors and magnitudes compatible with Tucana's RGB. Slits that would otherwise have remained empty (five of them) were allocated to random targets in the same magnitude range. We selected 50 objects distributed over two overlapping masks of 27 slits each; the observation of four objects were repeated on purpose for internal accuracy measurements. Targets thus selected covered an area up to the nominal King tidal radius of Tucana ($R_{\rm tidal}=3.7\arcmin$), as can be seen in Fig.~\ref{Fig:Targ}. 
	
	The adopted instrumental set-up and observing strategy were the same as in our previous studies \citep{Kacharov2017,Taibi2018,Hermosa-Munoz2019}, so we report only the essentials here.
	We used the 1028z+29 holographic grism together with the OG590+32 order separation filter in order to cover the wavelength range between 7700 and 9500\AA. 
	Slits had spatial sizes of $1\arcsec \times 10\arcsec$ ($8\arcsec$ in some cases to avoid overlaps) in the first mask (Tuc0) and of $1\arcsec \times 8\arcsec$ ($7\arcsec$ for overlaps) in the second one (Tuc1). This led to a binned spectral dispersion of 0.84 \AA\,pxl$^{-1}$ and a resolving power of $R = \lambda_{\rm cen}/\Delta\lambda \sim2600$ at $\lambda_{\rm cen}=8600$\AA\ (equivalent to a velocity resolution of 28 km\,s$^{-1}$\,pxl$^{-1}$). 
	Ten identical observing blocks (OBs\footnote{The observations are organized by ESO in blocks taking into account the time to actually spend on the object, including foreseen overheads. Data are later delivered as OB-datasets which include the scientific exposures related to the individual OBs together with the associated acquisition and standard calibration frames (biases, arc lamp, dome flat-fields).}) were taken for each pointing in order to reach the necessary S/N for velocity and metallicity measurements. In Table~\ref{table:2} we report a complete observing log. 
	
	The data were provided by ESO as individual OB-datasets, within the FORS2 standard delivery plan.
	We adopted the same data-reduction process as in \citet[][hereafter, T18]{Taibi2018}, based on IRAF\footnote{IRAF is the Image Reduction and Analysis Facility distributed by the National Optical Astronomy Observatories (NOAO) for the reduction and analysis of astronomical data: http://iraf.noao.edu/} routines and custom-made python scripts. Briefly, our pipeline was developed to organize and reduce each OB-dataset independently. After making bias and flat-field corrections on the two-dimensional (2D) multi-object scientific and lamp-calibration frames, these are also cleaned from cosmic rays and bad-rows. The 2D images are corrected for distortions in the spatial direction by rectifying their slit traces, in order to cut them into individual 2D spectra. The arc-lamp spectra are then used to find the wavelength solution to calibrate the scientific exposures, with a typical RMS accuracy of 0.05\AA. The wavelength calibration also has the important effect of rectifying the sky lines, which had been initially curved by the instrument disperser, helping to reduce the residuals during the sky-subtraction part. 
	The rectified wavelength-calibrated 2D individual scientific exposures are finally background subtracted, optimally extracted into 1D spectra and normalized by fitting the stellar continuum. 
	The median S/N around the CaT for the individual exposures was $\sim8$ \AA$^{-1}$.

	We then followed the approach already presented in T18 and \citet{Hermosa-Munoz2019} to stack together the repeated individual exposures for each target.
	To do so, we needed to account for possible zero-point displacements in the wavelength calibration, small slit-centering shifts and the different dates of observation. We used the numerous OH emission lines in the extracted sky background to refine the wavelength calibration of the individual spectra, using the IRAF \textit{fxcor} task to cross-correlate with a reference sky spectrum over the region $8250-9000$\AA. The calculated off-sets roughly varied between 5 and 25 km\,s$^{-1}$ with an average error of 2 km\,s$^{-1}$. 
	The correction for slit-centering shifts was done using the through-slit frames typically taken before each scientific exposure. The offsets were calculated as the difference in pixels between the slit center and the star centroid for every target per mask. The correction for each target was taken as the median value of all its slit-shifts; the associated error was the scaled median absolute deviation (MAD) of those values. We found slit-shifts in the range $\pm 0.1 - 9.5$ km\,s$^{-1}$ with errors of $\pm 1 - 5$ km\,s$^{-1}$. Finally, we obtained the heliocentric correction for all the individual exposures using the IRAF \textit{rvcorrect} task. All the above shifts were applied to the individual spectra using the IRAF \textit{dopcor} task. We then averaged together the repeated exposures of each target, weighting with their associated $\sigma$-spectra, given by the extraction procedure. Final error spectra were obtained accordingly. The median S/N around the CaT for the stacked spectra is $\sim27$ \AA$^{-1}$ (see also Table~\ref{table:P91} for the properties of each individual observed star). 
	
	\subsection{A new reduction for the Fraternali et al. (2009) FORS2 dataset} \label{subsec:data-F09}
	\citet[][hereafter, F09]{Fraternali2009} presented line-of-sight (l.o.s.) velocities and metallicities for 23 individual stars with magnitudes and colors compatible with Tucana's RGB (of which 17 were classified as members) from an earlier FORS2 MXU dataset (ESO program 69.B-0305(B), PI: E. Tolstoy). 
	We wished to use this catalog to increase our sample size; however, a comparison with the l.o.s. velocities and metallicities between the targets in common (3) with the F09 catalog showed significant systematic shifts in both quantities (see Appendix~\ref{subsec:vel-comparison} for details). For the sake of homogeneity, we therefore re-reduced the F09 dataset following the same procedure as adopted for the P91 sample. Hereafter, we refer to this additional sample as the P69 FORS2 dataset.
	
	The instrumental set-up of the P69 program was similar to that we adopted for our observations, with the difference that the size of each slit was of $1.2\arcsec \times 8\arcsec$, which translates into a slightly lower spectral resolution ($\approx30$ km\,s$^{-1}$\,pxl$^{-1}$). A total of 45 initial targets were assigned to an equivalent number of slits packed into a single pointing centered on Tucana. The total exposure time of the observations was 5.2~hours.
	
	The reduction of some slits, in particular the background subtraction step, turned out to be particularly problematic since these targets were not well centered along the slit spatial direction, but placed at their edges. This led to 15 extracted spectra with high sky-residuals, which made them unreliable for l.o.s. velocity and metallicity estimates. In addition, these targets had magnitudes and colors outside the RGB of Tucana. Therefore, we excluded them from the sample, alongside with two further objects whose extracted spectra did not show any CaT lines.
	
	The final P69 sample reduced to 28 objects, initially five more with respect to the published catalog of F09. Once stacked together, reliable spectra had a S/N $\sim22$ \AA$^{-1}$ (see also Table~\ref{table:P69} for the properties of each individual observed star). 
	
	Another issue was the lack of through-slit images in the provided data, although in F09 it is reported that the objects resulted well centered in the spectral direction after visual inspection during the spectroscopic run. However, we took into account the error related to the slit centering by assuming that it is a tenth of a pixel ($\sim3$ km\,s$^{-1}$, which is the typical error found checking for the slit centering) and adding it in quadrature during the velocity estimation step. 
	
	As it can be seen in Appendix~\ref{subsec:comparison-flames}, there is good agreement for the measurements of the stars in common between this dataset (the P69 FORS2) and the P91 FORS2 (and for those cases where there is disagreement, the source of it can be traced back).
	
	\begin{table*}
		\caption{Observing log of the P91 VLT/FORS2 MXU observations of RGB targets along the line-of-sight to the Tucana dSph. From left to right, column names indicate: the pointing field name; the field center coordinates; observing date and starting time of the scientific exposure; the exposure time in seconds; the starting airmass; the average DIMM seeing during the exposure in arcsec; the ESO OB fulfillment grades (a full description is reported in the notes below); the number of slits/observed objects per mask. For each field, the mask design remained identical in each OB.
			The total number of slits (54) is reported in the last row of the table. }             
		\label{table:2}      
		\centering          
		\begin{tabular}{c c c c c c c c}    
			\hline\hline
			Field & Position (RA, Dec)  & Date / Hour & Exp. & Airmass & DIMM Seeing & Grade\tablefootmark{$\star$} 
			& Slits\\ 
			& (J2000) & (UT) & (s) & & ($''$) & & \\
			\hline           
			Tuc0 & 22:41:56 -64:24:43	 & 2013-07-30 / 04:26 & 3400 & 1.44 & 0.78 & B & 27 (16+11)\\
			&  & 2013-08-01 / 04:46 & 3400 & 1.39 & 0.80 & A & \\ 
			&  & 2013-08-01 / 05:44 & 3400 & 1.32 & 0.92 & A & \\
			&  & 2013-08-01 / 06:42 & 3400 & 1.30 & 0.90 & A & \\  
			&  & 2013-08-05 / 06:04 & 3400 & 1.30 & 0.98 & A & \\  
			&  & 2013-08-05 / 07:02 & 3400 & 1.31 & 1.13 & A & \\  
			&  & 2013-08-13 / 03:49 & 3400 & 1.41 & 0.69 & A & \\ 
			&  & 2013-08-13 / 04:47 & 3400 & 1.33 & 0.57 & A & \\  
			&  & 2013-08-28 / 03:18 & 3400 & 1.36 & 0.65 & A & \\  
			&  & 2013-08-30 / 03:43 & 3400 & 1.33 & 0.89 & A & \\  
			Tuc1 & 22:42:04, -64:24:47 & 2013-09-05 / 02:07 & 3500 & 1.43 & 1.62 & C\tablefootmark{a}
			& 27 (20+7)\\
			& & 2013-09-06 / 03:34 & 3500 & 1.31 & 0.77 & A & \\ 
			& & 2013-09-06 / 04:37 & 3500 & 1.30 & 0.89 & A & \\  
			& & 2013-09-13 / 01:53 & 3500 & 1.40 & 1.49 & C\tablefootmark{b} 
			& \\ 
			& & 2013-09-13 / 02:54 & 3500 & 1.32 & 1.45 & B & \\ 
			& & 2013-10-01 / 01:35 & 3500 & 1.33 & 0.68 & A & \\
			& & 2013-10-25 / 02:17 & 3500 & 1.33 & 1.04 & B & \\  
			& & 2014-07-21 / 07:28 & 3500 & 1.30 & 0.92 & A & \\  
			& & 2014-07-21 / 08:40 & 3500 & 1.33 & 1.05 & A & \\  
			& & 2014-07-27 / 06:07 & 3500 & 1.32 & 0.70 & A & \\ 
			& & 2014-07-27 / 07:10 & 3500 & 1.30 & 0.60 & A & \\ 
			& & 2014-07-27 / 08:18 & 200    & 1.33 & 0.65 & C\tablefootmark{c} 
			& \\ 
			& & 2014-07-29 / 05:14 & 3500 & 1.37 & 0.62 & A & \\		
			\hline
			Total & &  &  &  &  & & 54 \\  
			\hline    
		\end{tabular}
		\tablefoot{\tablefoottext{$\star$}{ESO OB fulfillment Grades:
				A) Fully within constraints -- OB completed;
				B) Mostly within constraints, some constraint is ~10\% violated -- OB completed;
				C) Out of constraints -- OB must be repeated:}
			\tablefoottext{a}{Seeing increased up to $\sim1.2''$ at OB's end - FWHM of spectra $\sim1.0''$.}
			\tablefoottext{b}{Seeing, FLI and Moon distance out of constraints.}
			\tablefoottext{c}{Aborted after 200s because sky condition changed to thin clouds.}
		}
	\end{table*}
	
	\subsection{The FLAMES dataset} \label{subsec:data-flames}
	
	Recently, \citet[][hereafter, G19]{Gregory2019} have presented an additional sample of l.o.s. velocities (not metallicities) for individual stars in the direction of Tucana, taken with the FLAMES/GIRAFFE instrument at the VLT, as part of the ESO program 095.B-0133(A), PI: M. Collins. The spectrograph was used in MEDUSA mode, i.e. in multi-fiber configuration which allows for the simultaneous observation of up to 132 separate targets (sky fibers included). The instrument field of view (FoV) has a 25 arcmin diameter and each fiber has an aperture on the sky of 1.2 arcsec. The grating used was the LR8, centered on 8817 \AA\ and covering the CaT wavelength region, yielding a spectral resolution of $R\sim 6500$. 
	
	The authors reported the detection of 36 probable member stars, out to very large distances from Tucana's center, i.e. approximately to 10 half-light radii. Given the larger spatial region probed by these data with respect to the FORS2 P91 and P69 datasets (compare Fig.~\ref{Fig:Targ} of this work to Fig.~1 in G19), it was of interest to explore whether the G19 catalog of l.o.s. velocities could be used together with our determinations from the P91 and P69 FORS2 data. The comparison of the 6 stars in common between the P91 and G19 samples yields a discrepancy of $\sim$30 km\,s$^{-1}$ for 5 of the 6 stars and of about -150 km\,s$^{-1}$ for the other object, which does not allow to directly combine the measurements. G19 also report an offset of $\sim$23 km\,s$^{-1}$ between their velocities and those in the F09 catalog. We note that the offset of $\sim$30 km\,s$^{-1}$ is compatible with the 23 km\,s$^{-1}$ offset between G19 and F09 and the $\sim$7 km\,s$^{-1}$ offset we found when comparing our P91 velocities to the F09 catalog (see Appendix~\ref{subsec:vel-comparison} for further details). Given the above, we proceeded to perform our own reduction of the FLAMES/GIRAFFE data, the characteristics of which we briefly describe below. 
	
	The observations were taken on 6 nights spread between June and September 2015, using two different fiber setups covering the same area: the first one (Tuc-1) had a total of 7h exposure time accumulated over 7 OBs and the second setup (Tuc-2) got 6h exposure time taken within 6 OBs, of which one was repeated twice. Each OB consisted of $3\times1200 $sec exposures.
	The total number of individual targets was 164.
	The first setup had 14 and the second 15 fibers allocated to empty sky regions distributed over the entire FoV.
	Only few targets are spatially found inside the tidal radius of Tucana ($\sim$30 targets), with the others scattered over an area much larger than the nominal extension of the galaxy. In fact, the two pointings are off-centered by $\sim 9 \arcmin$ from the optical center of Tucana.
	
	The FLAMES data were downloaded from the science portal of the ESO archive as already processed spectra, i.e. pre-reduced, wavelength calibrated, extracted, corrected to the barycentric velocity, but with no sky subtraction applied. For each OB, ESO delivers spectra stacked at the OB-level for the science targets and several auxiliary data, including the individual scientific exposures within an OB for both scientific targets and sky fibers \footnote{See \url{http://www.eso.org/observing/dfo/quality/PHOENIX/GIRAFFE/processing.html} for details.}. 
	The only steps left to do were to perform the sky subtraction, combine the spectra having repeated exposures and finally calculate the radial velocities.
	
	We first verified the quality of the wavelength calibration by cross-correlating the sky lines of the scientific exposures with a template sky spectrum, placed at the rest-frame and obtained with a similar observational set-up \citep{Battaglia2011} \footnote{We further checked the wavelength calibration of the template spectrum itself using a high-resolution atlas of sky emission lines taken with the VLT/UVES instrument \citep{Hanuschik2003}, degraded to the spectral resolution of the template spectrum. The cross-correlation between these two spectra showed no significant wavelength shift.}. However, the delivered spectra are shifted to the heliocentric frame, including the sky lines, thus we had to remove this correction first. We performed then the cross-correlation for each spectrum of each OB-dataset using the IRAF \textit{fxcor} task, obtaining median offsets around 0.6 km\,s$^{-1}$ with a global scatter of 0.7 km\,s$^{-1}$.
	Therefore the uncertainty related to the wavelength calibration resulted well below those from the velocity measurement (as shown later).
	
	For the sky subtraction, we used the ESO \textit{skycorr} tool \citep{Noll2014}. The idea behind this code is to adopt a physically motivated group scaling of the sky emission lines with respect to a reference sky spectrum according to their expected variability given the date of the observations. We created the reference sky spectrum by first median combining the spectra of the fibers allocated to sky within the individual sub-exposures of each OB, and then by median combining the results for the individual sub-exposures. The optimized line groups in the reference sky spectrum are scaled to fit the emission lines in the science spectra and finally subtracted together with the sky continuum. Error spectra are also an output of the code. We used default input parameter while running \textit{skycorr}. This method yielded satisfactory results for the majority of stars, particularly for spectra with low S/N. 
	
	The sky-subtracted science spectra were then normalized using a Chebyshev polynomial of order 3, together with their error spectra. 
	Finally, repeated exposure of individual targets (including those in common between the Tuc-1 and Tuc-2 set-ups) were stacked together using a weighted average, with their error spectra combined accordingly.
	The typical S/N was $\sim$11 \AA$^{-1}$, although in some cases it was as low as $\sim$2 \AA$^{-1}$ (see also Table~\ref{table:FLAMES} for the properties of each individual observed star). 
	
	The average S/N measured in our reduction of the FLAMES/GIRAFFE spectra is in good agreement with the S/N obtained from the GIRAFFE Exposure Time Calculator (ETC) considering typical values from the observed dataset: we used a black body template of $T_{\rm eff}=4500$ K, an \textit{I}-band magnitude of 21, an airmass of 1.2, a moon illumination fraction of 0.2, a seeing of 1.0\arcsec,  an object-fiber displacement  of 0.3\arcsec and a total exposure time of 10 hours (36000 sec). With this setting we obtained a calculated S/N of 15\,\AA$^{-1}$, close to our typical S/N value. 
	
	\subsection{Photometric data}
	Photometric data were used for all the three datasets (FORS2 P91 and P69, and FLAMES/GIRAFFE) to exclude those objects whose magnitude and color were not compatible with being stars on Tucana's RGB (see Sect.~\ref{sec:membership}), and for the FORS2 datasets to determine metallicities from the equivalent width of the CaT lines using calibrations from the literature (see Sect.~\ref{sec:measurements}). 
	
	We used the pre-imaging FORS2 photometric catalog introduced in Sect.\ref{subsec:data-fors2} to associate \textit{V}- and \textit{I}-band magnitudes to the target stars in the P91 and P69 spectroscopic datasets. The photometric catalog was astrometrized and the instrumental magnitudes calibrated taking as reference the publicly available catalog from \citet{Holtzman2006} obtained with the Wide Field and Planetary Camera 2 of the Hubble Space Telescope (HST/WFPC2). We used the aperture-photometry catalog provided in the Johnson's \textit{UBVRI}-system, to actually find the astrometric solution of the FORS2 catalog and to perform the photometric calibration using the suite of codes \textit{CataXcorr} and \textit{CataComb}, kindly provided to us by P. Montegriffo and M. Bellazzini (INAF-OAS). 
	
	The case of the FLAMES dataset, on the other hand, was different. Since we did not have a photometric catalog covering an area as wide as that of the spectroscopic targets, we used instead the publicly available photometry of individual point-sources from the first data release of the Dark Energy Survey (DES-DR1, \citealp{Abbott2018}). We found a match for 154 out of 164 spectroscopic targets, considering a tolerance radius of 1 arcsec. To be conservative, we did not exclude from the photometric selection those targets that did not have a match in the DES-DR1 photometry. The DES-DR1 \textit{griz}-photometry needed then to be converted first to the SDSS \textit{griz}-system\footnote{https://des.ncsa.illinois.edu/releases/dr1/dr1-faq} and finally to the Johnson's system (see \citealp{Jordi2005}).
	The FLAMES data resulted to have magnitudes as high as $V\sim23$, which was also the limit of the DES-DR1 catalog and, consequently, some targets had relatively large magnitude errors associated ($\delta_{\rm mag} \sim 0.2$).
	
	\section{L.o.s. velocity and metallicity measurements} \label{sec:measurements}
	
	The determination of line-of-sight (l.o.s.) velocities and metallicities ([Fe/H]) from the stacked spectra was done as in T18. 
	While we could measure both l.o.s. velocities and metallicity for the FORS2 P91 and P69 data, only l.o.s. velocities could be determined for FLAMES/GIRAFFE data due to the low S/N ratio of those spectra.
	
	L.o.s. velocities were obtained using the \textit{fxcor} task by cross-correlating with a synthetic spectrum resembling a low-metallicity RGB star convolved at the same spectral resolution of the dataset under consideration. For FORS2 we used the same 
	template as in T18, and cross-correlated in the wavelength range $8400-8700$\AA. 
	For the FLAMES dataset we used a template from \citet{Zoccali2014}, that is a synthetic spectrum of a star with $T_{\rm eff}=4750$ K, log$(g)=2.5$ and [Fe/H] $=-1.3$ dex. In this case we used the region around the two reddest lines of the CaT for the cross-correlation, since the first line often suffered of high residuals left by the subtraction of a sky emission line. This was not necessary for the FORS2 spectra that suffer less from this problem due to their higher S/N.
	
	In the following, we only keep objects whose stacked FORS2 and FLAMES spectra have a S/N$\gtrsim10$ \AA$^{-1}$, since this is the limit where the velocity errors provided by \textit{fxcor} task appear reliable (see tests in Appendix~\ref{subsec:sanity-vel}). Average velocity errors for the FORS2 (FLAMES) spectra resulted to be $\sim$7 km\,s$^{-1}$ ($\sim$9 km\,s$^{-1}$).
	
\begin{figure*}
		\centering
		\includegraphics[width=\hsize]{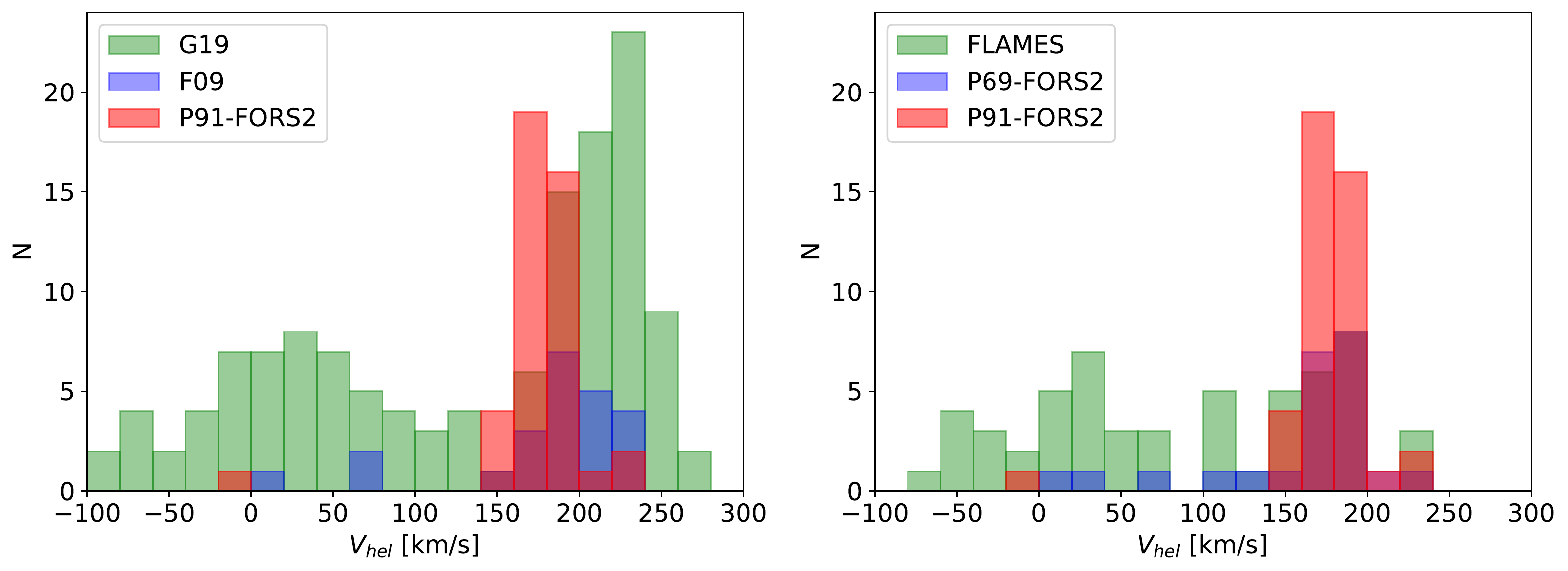}
		\caption{Histogram of the l.o.s. velocity measurements from the analyzed datasets. \textit{Left:} comparing the velocities from the P91-FORS2 dataset with those from literature, i.e. \citet{Fraternali2009} and \citet{Gregory2019}. \textit{Right:} same comparison using, however, the velocity measurements from our reduction of the P69-FORS2 and FLAMES datasets. Note that, in the left panel, the peaks of the histograms fall at different velocities, while in the right panel, where the datasets shown have been analyzed in a homogeneous way, these differences are absent.
		}
		\label{Fig:flames-hist}
	\end{figure*}

	As can be seen from the histogram in the right panel Fig.~\ref{Fig:flames-hist}, the three datasets show a clear peak in the l.o.s. velocities around 180 km\,s$^{-1}$, where we roughly expect to find the members of Tucana. The FLAMES dataset presents a higher fraction of contaminants due to the large area covered by the observations. 
	With our homogeneous analysis we show a clear detection of the stars with velocities compatible with Tucana in all three datasets and no significant offsets, unlike what we found in direct comparison of our velocity measurements with those published in F09 and G19, as shown in the left panel of Fig.~\ref{Fig:flames-hist}. We refer the reader to Appendix~\ref{subsec:vel-comparison} and Sect.~\ref{subsec:kin-comparison} for a detailed comparison with the studies from literature.
	
	To estimate the [Fe/H] values we adopted the \citet{Starkenburg2010} relation, which is a function of the equivalent widths (EWs) of the two reddest CaT lines, and of the ($V-V_{\rm HB}$) term, where $V_{\rm HB}$ is the mean magnitude of the galaxy's horizontal branch (HB). We obtained the EWs from the continuum normalized stacked spectra by fitting a Voigt profile over a window of 15 \AA\ around the CaT lines of interest and using in the fitting process the corresponding error-spectra as the flux uncertainty at each pixel. The errors on the EWs were then calculated from the covariance matrix of the fitting parameters.
	For the $V_{\rm HB}$ we adopted the value of 25.32 from \citet{Bernard2009}. 
	Uncertainties on the [Fe/H] values were obtained by propagating the errors on the EWs accordingly. Typical [Fe/H] errors were found to be around $0.15 - 0.25$ dex.
	
	\section{Membership \& kinematic analysis} \label{sec:membership}
	
	Before proceeding with the analysis of Tucana's kinematic and chemical properties, we need to identify the stars that are probable members of Tucana and weed out possible contaminants (foreground MW stars and background galaxies). We followed the same steps for the membership selection in all the catalogs we analyzed. 
	
	We first selected targets located approximately along the RGB of Tucana making a selection in magnitude and color.
	We used a set of isochrones \citep{Girardi2000,Bressan2012} with age $t_{\rm age}=12.6$ Gyr and [Fe/H] $\sim -2.3$ dex, and age $t_{\rm age}=8$ Gyr and [Fe/H] $\sim -0.4$ dex to fix the blue and red color limits on the color-magnitude diagram (CMD). This color range was chosen to broadly cover the expected range of metallicities and stellar ages obtained from the SFH analysis of Tucana \citep{Monelli2010,Savino2019}. For the FLAMES dataset, due to the larger errors in the associated photometric dataset, we broadened by 0.2 mags the blue and red color limits applied to the FORS2 case. 	
	We further excluded from all catalogs the targets with spectra that showed high sky-residuals (and with S/N values $<10$ \AA$^{-1}$, as previously discussed). Our P91 sample reduced from 50 to 43 targets, the P69 one from 28 to 23 objects, and the FLAMES one from 164 to 58.
	
	Given that the target selection of the three datasets was carried out in completely independent ways, and might therefore include different biases, we proceed by determining memberships and kinematic parameters by considering first our P91 FORS2 dataset on its own, a second time by combining it to the P69 sample, and finally considering all the three sets. When we found common targets among the combined catalogs, we first kept those from the FORS2 datasets, with a higher priority for those of P91. Therefore, the combined FORS2 dataset has 63 targets while including the FLAMES data added to 109.
	
	We continued then by assigning a membership probability to the individual targets in each considered dataset, applying a method based on the expectation maximization technique outlined in \citet{Walker2009}, but with the few modifications introduced by \citet{Cicuendez2018}. 
	Briefly, this approach allows to carry out a Bayesian analysis to obtain the kinematic parameters of interest while assigning a probability of membership $P_{\rm M_i}$ to each i-star by maximizing the following log-likelihood equation:
	\begin{equation}
	{\rm ln}\, L = \Sigma_i P_{\rm M_i} {\rm ln}\left [   P_{\rm mem}\,P_{\rm rad} \right ] + \Sigma_i (1-P_{\rm M_i}) {\rm ln}\left [   P_{\rm non}\,(1-P_{\rm rad}) \right ] 
	\end{equation} 
	where $P_{\rm mem}$ is the targets probability distribution depending on their l.o.s. velocities, $P_{\rm rad}$  and $P_{\rm non}$ are the prior probabilities related to the surface density profile of the galaxy and the presence of possible contaminants, respectively, while $P_{\rm M_i}$ is defined as:
	\begin{equation}
	P_{M_i} = \frac{ P_{\rm mem}\,P_{\rm rad}}{ P_{\rm mem}\,P_{\rm rad} +  P_{\rm non}\,(1-P_{\rm rad})}
	\end{equation} 
	Both equation were adapted from Eq.~(3) and (4) of \citet{Walker2009}, respectively. 
	
	We run the Bayesian analysis using the \textit{MultiNest} code \citep{Feroz2009, Buchner2014}, a multi-modal nested sampling algorithm, in order to obtain the kinematic parameters and the membership probabilities at once, as in \citet{Cicuendez2018}. A further output of this code is the Bayesian evidence, which gives us the possibility to compare different kinematic models according to their statistical significance.
	
	The spatial prior probability as a function of radius $P_{\rm rad}(R_i)$ accounts for the fact that it is more probable to observe a member star near the galaxy's center than in its outer regions. To this aim we assumed an exponentially decreasing surface number density profile, which takes into account a uniform background surface number density. 
	The parameters of the profile were obtained from a VLT/VIMOS photometric dataset centered on Tucana, kindly provided by G. Beccari (ESO) and M. Bellazzini (INAF-OAS). This photometry was preferred to the FORS2 pre-imaging and the DES-DR1 photometry used for the CMD-selection since it is much deeper (almost 3 magnitudes in \textit{I}-band), although it extends up to the tidal radius. 
	To obtain the best-fitting structural parameters we applied a Bayesian Monte Carlo Markov chain (MCMC) analysis following the density profile of the RGB stars, as done in \citet{Cicuendez2018}, while accounting for contamination.
	The assumed profile resulted to be a good representation of the observed surface number density profile for Tucana and the best-fitting structural parameters (see Table~\ref{table:struct-params}) are perfectly compatible with the parameters reported by \citet{Saviane1996}\footnote{Also these authors performed an exponential fit to the RGB density profile which, however, was obtained from a shallower photometry covering approximately the same area as that of VLT/VIMOS.}.
	
	The prior probability of contamination by foreground stars $P_{\rm non}$ was based on the Besan\c{c}on model \citep{Robin2003}. The generated distribution of l.o.s. velocities was well fitted by a Gaussian profile ($\bar{v}_{\rm Bes}=57$ km\,s$^{-1}$; $\sigma_{\rm Bes}=99$ km\,s$^{-1}$).
	The contamination model was generated in the direction of Tucana over an area equivalent to a FLAMES/GIRAFFE pointing selecting stars over the range of colors and magnitudes described above. 
	
	The l.o.s. velocity distribution of the probable member stars $P_{\rm mem}(v_{\rm i})$ was assumed to be Gaussian, as in T18, and accounted for the different kinematic models according to the following rotational term: $v_{\rm rot}(R_i)\,{\rm cos}(\theta - \theta_i)$, with $R_i$ being the angular distance from the galaxy's center, $\theta_i$ the position angle (measured from north to east) of the i-th target star, $\theta$ the position angle of the kinematic major axis (i.e. the direction of the velocity gradient, perpendicular to the axis of rotation), while $v_{\rm rot}(R_i)$ is the modeled rotational velocity term. We fit and compared three kinematic models: a dispersion-only model (i.e. with the velocity rotation term set to zero), a model with rotational velocity linearly increasing with radius ($v_{\rm rot}(R_i) = k\,R_i$) and a flat one ($v_{\rm rot}(R_i) = v_{\rm c} = $ constant). 	
	
	The free kinematic parameters were the following ones: the systemic velocity $\bar{v}_{\rm hel}$ and velocity dispersion $\sigma_{\rm v}$ common to the three models, the position angle $\theta$ of the kinematic major axis, the velocity gradient \textit{k} of the linear rotation model, and the constant rotational velocity $v_c$ of the flat model. 
	In our definition, the position angle $\theta$ varies between 0$^\circ$ and 180$^\circ$, which means that a rotation signal (either expressed as \textit{k} or $v_{\rm c}$) having a negative sign implies a receding velocity on the West side of the galaxy (and would be equivalent to a positive gradient adding 180$^\circ$ to $\theta$).
	The model evidences, \textit{Z}, were combined together through the Bayes factor, $B_{1,2}=Z_1/Z_2$, where the sub-scripts indicate a given model. 
	To quantify the statistical significance of one model with respect to another we made use of the Jeffrey's scale, based on the natural logarithm of the Bayes factor: positive values of $(0-1), (1-2.5), (2.5-5), (5+)$ corresponds to inconclusive, weak, moderate and strong evidence favoring one model over the other (T18; \citealp{Hermosa-Munoz2019}, but also \citealp{Wheeler2017}). In our case we had the following Bayes factors: ln~$B_{\rm lin,flat}$, comparing the evidences of the two rotational models, and ln~$B_{\rm rot,disp}$, between the evidences of the best rotational model (choosing the one that has the largest \textit{Z}) and that of the dispersion-only one.
	
	We used the following priors for the kinematic parameters: $-50 < (\bar{v}_{\rm hel} - v_g)\,[{\rm km\,s^{-1}}] < 50$, where $v_g$ is the initial mean value of the velocity distribution, $0 < \sigma_{\rm v}\,[{\rm km\,s^{-1}}] < 50$, $-50 < k\,[{\rm km\,s^{-1}\,arcmin^{-1}}] < 50$, and $-50 < v_c \,[{\rm km\,s^{-1}}] < 50$.  The prior over $\theta$ was set iteratively: we initially chose the prior range $0 < \theta < \pi$, run the MultiNest code a first time in order to obtain the maximum value $\theta_m$ from the $\theta$ posterior distribution and run again the MultiNest code updating the prior range to  $- \pi/2 < \theta -  \theta_m < +\pi/2$. This choice accounted for the limit case of $\theta$ near $0$ or $\pi$.
	
	Results of the recovered probability-weighted kinematic parameters and evidences for the three analyzed datasets, together with the effective number of probable members (defined as $N_{\rm eff}\approx \Sigma_{\rm i} P_{\rm M_i}$) are reported in Table~\ref{table:multinest}. 
	The $P_{\rm M_i}$ finally assigned to each target were those obtained from the most significant kinematic model. 
	
	The P91 FORS2 dataset yields 39 effective members, while the inclusion of the P69 and FLAMES data adds about 15 more effective members. These numbers already double the member stars reported by F09 (17) and G19 (36, although when accounting for their probability of membership they reduce to $\sim20$ effective members).
	We shall note that for each analyzed dataset, the effective number of members resulted to be approximately the same for each kinematic model.
	In all cases, the systemic velocity is stable around 180 km\,s$^{-1}$, while the velocity dispersion converges to 6 km\,s$^{-1}$ for the dispersion-only model, which resulted to be the most statistically significant one. Both the systemic velocity and velocity dispersion values are significantly lower than what reported in literature studies.
	We refer to Sect.~\ref{sec:kinematics} for the full discussion of the results from the kinematic analysis.
	
	We should also note that the adopted method does not take into account the different selection functions of the various datasets, that were independently built. Therefore, to test whether this introduced a bias in our results, we repeated the analysis relaxing the assumption of an exponentially declining surface number density profile for Tucana's stars and simply required it to be monotonically decreasing \citep[like in][]{Walker2009}. No significant difference in the results was found.

	\section{Kinematic results} \label{sec:kinematics}
			
	\begin{table*}
		\caption{Parameters and evidences resulting from the probability-weighted Bayesian kinematic analysis for all the datasets analyzed in this work. The reported values of the kinematic parameters represent the median of the corresponding marginalized posterior distributions, with 1-$\sigma$ errors set as the confidence intervals around the central value enclosing $68 \%$ of each distributions.}
		\label{table:multinest}
		\centering          
		\begin{tabular}{c c c c c c c c c|l}    
			\hline\hline
			Sample & N$_{\rm in}$ & N$_{\rm eff}$ & Model & $\bar{v}_{\rm hel}$ & $\sigma_{\rm v}$ & \textit{k} & $v_{\rm c}$ & $\theta$ & Bayes factor \\ 
			 &  &  &  & (km\,s$^{-1}$)   &  (km\,s$^{-1}$)   &   (km\,s$^{-1}$\,arcmin$^{-1}$)   &   (km\,s$^{-1}$)   &  (deg) & \\
			\hline     
			& & & Linear   & 178.9$^{+1.4}_{-1.5}$ &  6.2$^{+1.9}_{-1.7}$ & $-$1.9$^{+2.4}_{-5.1}$ &  & 5$^{+40}_{-38}$ & ln$B_{\rm lin,flat}=0.4$ \\
			P91 & 43 & 39 & Flat  & 179.0$^{+1.4}_{-1.3}$ & 6.0$^{+2.0}_{-1.6}$ &  & 0.5$^{+2.3}_{-2.1}$ & 164$^{+50}_{-64}$ & ln$B_{\rm rot,disp}=-2.6$ \\
			& & & No rotation & 179.0$^{+1.3}_{-1.3}$ & 5.7$^{+1.7}_{-1.5}$ &  &  &  & \\           
			\hline\hline
			& & & Linear  & 180.0$^{+1.4}_{-1.4}$ & 7.4$^{+2.3}_{-1.7}$ & 3.2$^{+5.9}_{-2.9}$ &  & 173$^{+23}_{-38}$ & ln$B_{\rm lin,flat}=0.6$ \\
			P91$+$P69 & 63 & 53 & Flat  & 180.0$^{+1.3}_{-1.4}$ & 6.6$^{+1.7}_{-1.3}$ &  & 1.0$^{+2.2}_{-2.0}$ & 152$^{+48}_{-62}$ & ln$B_{\rm rot,disp}=-2.3$ \\
			& & & No rotation & 180.0$^{+1.3}_{-1.2}$ & 6.4$^{+1.6}_{-1.2}$ &  &  &  & \\           
			\hline     		     
			&	&	& Linear  & 180.2$^{+1.5}_{-1.5}$ & 8.4$^{+2.2}_{-2.0}$ & 6.3$^{+4.5}_{-4.5}$ &  & 167$^{+16}_{-26}$ & ln$B_{\rm lin,flat}=1.6$ \\
			All comb. & 109 & 55 & Flat & 179.9$^{+1.3}_{-1.4}$ & 7.0$^{+1.9}_{-1.4}$ &  & 1.5$^{+2.2}_{-2.2}$ & 140$^{+47}_{-61}$ & ln$B_{\rm rot,disp}=-1.1$ \\
			&	&	& No rotation & 180.1$^{+1.2}_{-1.3}$ & 6.6$^{+1.6}_{-1.2}$ &  &  &  & \\           
			\hline
		\end{tabular}
	\end{table*}
	
	\begin{figure*}
		\centering
		\includegraphics[width=\hsize]{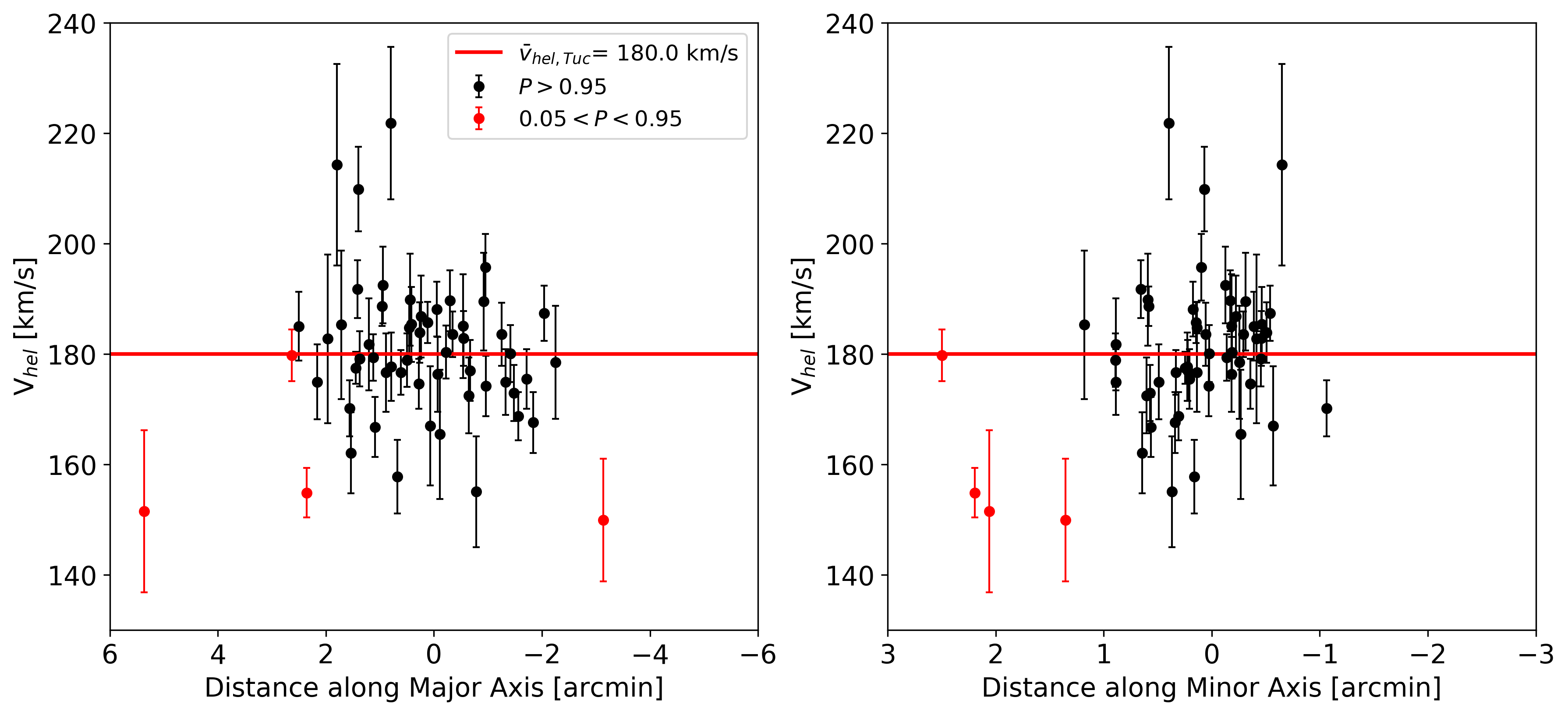}
		\caption{Line-of-sight velocity distributions of the stars from the combined FORS2 + FLAMES dataset, having membership probabilities $P>0.95$ (in black) and $0.05<P<0.95$ (in red). The red solid line indicates the systemic velocity obtained from the kinematic analysis. \textit{Left panel}: the distribution along the optical major axis; \textit{right panel}: along the minor axis.} 
		\label{Fig:kin1.1}
	\end{figure*}
	
	The analyses of the properties of the stellar component of Tucana carried out in the literature indicated a system with a relatively high l.o.s. velocity dispersion ($\sigma_v\sim15$ km\,s$^{-1}$) compared to other similarly luminous companions (see e.g. the compilation for LG dwarfs of \citealp{Kirby2014} and \citealp{Wheeler2017}). The works of F09 and G19 also reported the tentative presence of a velocity gradient likely due to internal rotation, since perspective effects related to transverse motion are negligible at the distance of Tucana. However, on a Bayesian analysis of the rotational support of the stellar component of LG galaxies using data from the literature, \citet{Wheeler2017} found no significant evidences for rotation and a l.o.s. velocity dispersion as high as 21 km\,s$^{-1}$, when analyzing the F09 catalog. Finally, there seems to be not much agreement about the systemic velocity of Tucana among F09 and G19, showing an offset of $\sim20$ km\,s$^{-1}$ at more than 3-$\sigma$ significance.
	
	As shown in Table~\ref{table:multinest}, our results are remarkably similar for all the cases we analyzed: i.e. when using the P91 FORS2 data alone, combining P91 and P69 or with all the datasets together: the systemic velocity and the velocity dispersion settled around $\sim$180 km\,s$^{-1}$ and 6 km\,s$^{-1}$, respectively. Furthermore we found no significant evidence of rotation, with the dispersion-only model moderately favored in all of the cases. We shall note a slight increase of the velocity dispersion for the rotational models: this is caused by few targets acquiring a higher membership probability due to the different fitted models. However, the effect is small and all the velocity dispersion values we obtained are compatible within 1-$\sigma$.
	
	We further performed a simpler analysis on the three datasets, considering only those targets with the highest membership probabilities (i.e. having $P_{\rm M}>0.95$) looking for the kinematic parameters of a dispersion-only model, whose results confirmed those of the probability-weighted analysis. Performing the same test by adding those stars with a lower membership probability (i.e. having $0.05<P_{\rm M}<0.95$), would instead increase the velocity dispersion up to 8 km\,s$^{-1}$, which is still within 1-$\sigma$ from the previous results taking into account the error bars. We also note that these stars are found further away from the center of Tucana compared to the more probable members (see Fig.~\ref{Fig:kin1.1}), but are still inside the tidal radius of the galaxy. 
	However, it is hard to discern whether the increase in the velocity dispersion seen when including stars with lower membership probability is caused by a radially increasing velocity dispersion profile, or simply because (all or part of) these stars are contaminants. 
	Considering those stars with a lower membership probability that have metallicity measurements, they seem to be preferentially metal-poor (see Fig.\ref{Fig:met1.2}). Therefore, it may also be that the increase in $\sigma_v$ could be caused by the preferential inclusion of metal-poor stars with a hotter velocity dispersion than the more metal-rich stars \citep[][]{Tolstoy2004,Battaglia2006,Battaglia2008b,Battaglia2011,Amorisco+Evans2012}. We refer to Sect.\ref{subsec:met-multipops} for more details on this point.
	
    One should caution that underestimated (optimistic) or overestimated (pessimistic) velocity errors may impact on the measured velocity dispersion. We refer the reader to Fig.~1 in \citet{Koposov2011} for an analysis of the impact of underestimated/overestimated velocity errors as a function of the ratio between the true error and the true velocity dispersion. However, as reported in Appendix~\ref{sec:A}, we have conducted several consistency tests where we have showed that the velocity errors are well determined.
    
	Therefore, our results for Tucana point to a value for the velocity dispersion which is very unlikely to exceed 10 km\,s$^{-1}$. 
	We assume as our reference value $\sigma_{\rm v} = 6.2_{-1.3}^{+1.6}$ km\,s$^{-1}$, averaging between the results of the dispersion-only model from the analyzed datasets.
	
	\subsection{MultiNest mock tests} \label{subsec:kin-mock}
	
	Although we have found that the rotation signal in our catalog is not statistically significant, we have conducted a series of mock tests in order to explore which rotational properties can be detected according to the characteristics of our data.
	To this aim, we followed the approach already introduced in T18 and \citet{Hermosa-Munoz2019}: we produced mock catalogs of l.o.s. velocities assuming the same number, spatial position, velocity distribution and velocity uncertainties of the observed data.  Our base catalog was the combined FORS2 P91 + P69 + FLAMES dataset after applying a $P_{\rm M}>0.05$ cut; the inclusion of less probable members was a compromise to have the highest number of targets (57) within the largest spatial area.
	To each target we assigned a mock velocity $v_{\rm mock}$ randomly extracted from a Gaussian distribution centered on zero and of standard deviation equal to the assumed velocity dispersion, fixed at $\sigma_{\rm v,mock}=6$ km\,s$^{-1}$. This value of $\sigma_{\rm v,mock}$ was set according to the converging results from the probability-weighted analysis of our datasets.
	We further added a projected linear rotational component $v_{\rm rot, mock}$, such that $v_{\rm rot, mock}$/$\sigma_{\rm v, mock} = n =  1.5, 1.0, 0.75, 0.5, 0.25, 0$ at the half-light radius. These correspond to velocity gradients of $k = 11.2, 7.5, 5.6, 3.7, 1.9, 0$ km\,s$^{-1}$\,arcmin$^{-1}$, that we simulated at three different position angles, starting from the P.A. of the optical semi-major axis (97$^\circ$) and then adding 45 and 90 degrees (optical semi-minor axis). 
	We chose an underlying linear rotation component since it resulted in higher evidence compared to a flat rotation model when analyzing our data.
	Each case was simulated $N = 1000$ times, in which we run our Bayesian kinematic analysis fitting just the linear rotation and the dispersion-only models in order to recover the related parameters and evidences. Results are showed in Table~\ref{table:mock}. 
	
	Results from the tests indicate that the linear rotation would be spotted with high significance for velocity gradient values $\geq 5.6$ km\,s$^{-1}$\,arcmin$^{-1}$ aligned with the projected optical major axis.
	If the underlying rotation instead is milder, e.g. with velocity gradients $\leq 3.7$ km\,s$^{-1}$\,arcmin$^{-1}$, the recovered evidences for rotation are weak, inconclusive or favoring the dispersion-only model, as we move through decreasing values of \textit{k} and through different position angles.
	In any case, it is evident that if Tucana has a weak rotation signal (i.e. $k\leq 3.7$ km\,s$^{-1}$\,arcmin$^{-1}$), with the data at our hand we could not detect it with high significance, in particular if it is not aligned with the optical major axis. 
	If instead Tucana has a velocity gradient as that reported in F09 and G19, i.e. with a value close to $k=5.6$ km\,s$^{-1}$\,arcmin$^{-1}$ along the major axis, with our data we would have detected it with high significance which, however, has not been the case. 
	
	Therefore, it seems that if rotation is actually present in Tucana, it probably is at a level of $v_{\rm rot}$/$\sigma_{\rm v} \lesssim 0.5$, and to detect it with high significance a better sampling would be needed, in particular at radii around $3\lesssim  R/R_e \lesssim 5$.
	
	We note that these results are conservative: in fact if we would have considered an input velocity dispersion as high as 10 km\,s$^{-1}$ the linear rotation signal would have been even more difficult to detect.
	
	\subsection{Comparison with other works} \label{subsec:kin-comparison}
	
	In Appendix~\ref{subsec:vel-comparison}, we perform a comparative analysis of the  velocity measurements for the individual stars derived in this work and those in F09 and G19. Here we focus instead on the comparison of the recovered velocity parameters from the kinematic analysis of the different works.
	
	First, we have run our code on the F09 velocity measurements for member stars, and the same for G19, in order to see if we were able to recover their results. Considering a dispersion-only model, we found a 1-$\sigma$ agreement between the recovered velocity parameters and those reported by these authors. Therefore, our procedure is not introducing a bias and we can directly compare with the reported values in F09 and G19.
	
	We found for the systemic velocity an offset of $\sim 15$ (35) km\,s$^{-1}$ between the values reported by F09 (G19) and us -- $\bar{v}_{\rm hel,F09}=194.0\pm4.3$ km\, s$^{-1}$ and $\bar{v}_{\rm hel,G19}=216.7^{+2.9}_{-2.8}$ km\, s$^{-1}$. These offsets are somewhat higher but still compatible with those reported in Appendix~\ref{subsec:vel-comparison}, so we refer the reader to that section for an analysis of the possible causes. We stress that, if we analyze our reduction of the P69 and the FLAMES data on their own, we obtain systemic velocities compatible at the 1-$\sigma$ level with our value of $\sim$180 km\, s$^{-1}$; therefore, the differences encountered appear to be related to the treatment of the datasets.
	
	On the other hand, the velocity dispersion values (without accounting for the presence of possible gradients) reported by F09 and G19 -- $\sigma_{\rm v, F09}=15.8^{+4.1}_{-3.1}$ km\,s$^{-1}$ and $\sigma_{\rm v, G19}=14.4^{+2.8}_{-2.3}$ km\,s$^{-1}$, differ from our reference $\sigma_{\rm v}$ value at almost the 3-$\sigma$ level. 
	
	If we were to analyze our reduction of the P69 FORS2 dataset alone, we would have obtained 17 effective members out of 23 input targets, showing a velocity dispersion value of $\sigma_{\rm v,P69}=11.1^{+3.7}_{-2.7}$ km\, s$^{-1}$, associated to a highly significant rotation signal. We noticed, however, that this gradient is driven by just two targets that have a very low membership probability when analyzing the combined FORS2 dataset. If we exclude them, the velocity dispersion would drop to $\sim$8 km\, s$^{-1}$ and the velocity gradient basically disappears. Therefore, it seems here that the low-number statistics strongly limits the conclusions we could get on the kinematic status of Tucana by the P69 dataset on its own.
	
	The comparison with G19 results are even more puzzling. If we were to apply the same exercises as for the P69 dataset, analyzing the FLAMES data on their own, we would have obtained 11 effective members out of 58 targets which, however, could not resolve the $\sigma_v$ value.  
	This is probably due to the combination of a small number of effective members and the fact that the average velocity error of stars with a high probability of membership ($\delta_{\rm v}\sim$6 km\, s$^{-1}$) is comparable to the $\sigma_v$ value we are finding in our main kinematic analysis.
	
	Furthermore, we want to underline that we found several differences when comparing our velocity measurements to those of G19, as described in Appendix~\ref{subsec:vel-comparison}. We suspect that the way the data were actually sky subtracted has led to an excess of stars with velocities around 220 km\,s$^{-1}$ in the G19 work, probably due to a combination of sky-line residuals and low S/N which would have created fake CaT features.

	\subsection{Implications for Tucana's dark matter halo properties} \label{subsec:kin-DM}
	
	As previously discussed, the analysis of the internal kinematic properties of Tucana yields consistent results across the combination of datasets. Our best value for the velocity dispersion of $\sigma_{\rm v} = 6.2_{-1.3}^{+1.6}$ km\,s$^{-1}$ is significantly lower ($\sim$3-$\sigma$) than what reported in the literature by both F09 and G19 (see also the discussion in the previous section), but closer now to the values observed for other similarly luminous dwarf galaxies of the Local Group \citep[see e.g.][]{Kirby2014, Revaz+Jablonka2018}. 
	
	We use the \citet{Wolf2010} mass-estimator valid for dispersion-supported spherical systems to calculate Tucana's dynamical mass inside the half-light radius,  $M_{1/2} = 3 G^{-1} \sigma_{\rm v}^{2} r_{1/2} \approx 4 G^{-1} \sigma_{\rm v}^{2} R_{\rm e}$, where \textit{G} is the gravitational constant and $r_{1/2}$ is the 3D de-projected half-light radius which can be approximated to $4/3 R_{\rm e}$. Using the values from Table~\ref{table:1} and substituting for $\sigma_{\rm v}$, we obtain  $M_{1/2}=0.7_{-0.3}^{+0.4}\times 10^7 M_\odot$, which corresponds to a mass-to-light ratio within the half-light radius of $M_{1/2}/L_V = 13_{-7}^{+8} \, M_\odot/L_\odot$, assuming a luminosity of $L_V = 5.5 \pm 1.5 \times 10^5 \, L_\odot$ (adapted from \citealp{Saviane1996}). 
	
	Using instead the velocity dispersion values from F09 and G19, applying the \citet{Wolf2010} mass-estimator we would obtain $M_{\rm 1/2, F09}=4.8^{+2.6}_{-2.0}\times 10^7 M_\odot$ and  $M_{\rm 1/2, G19}=3.9^{+1.6}_{-1.4}\times 10^7 M_\odot$ respectively, which are more than four time as large as our own best estimation of Tucana's mass.
	
	Our measurements for velocity dispersion and dynamical mass of Tucana provide a new perspective to the discussion found in the literature. 
	If we instead of $M_{1/2}$ we use the value of the circular velocity $V_{\rm circ}(r_{1/2}) = \sqrt{3} \sigma_{\rm v}$, we obtain for Tucana a value of $11_{-2}^{+3}$ km\,s$^{-1}$, assuming our reference value for $\sigma_{\rm v}$. This $V_{\rm circ}(r_{1/2})$ is comparable to those of other similarly luminous dwarfs like Carina, Sextans and Leo~II, as it is possible to see from Fig.~10 in G19, but also from Fig.~1 in \citet{Boylan-Kolchin2012}. From this last reference, we obtain that Tucana should live in a NFW dark matter halo having a maximum circular velocity $V_{\rm max}\leq24$ km\,s$^{-1}$, as is the case for many dSphs of the LG.
	
	\begin{figure}
		\centering
		\includegraphics[width=\hsize]{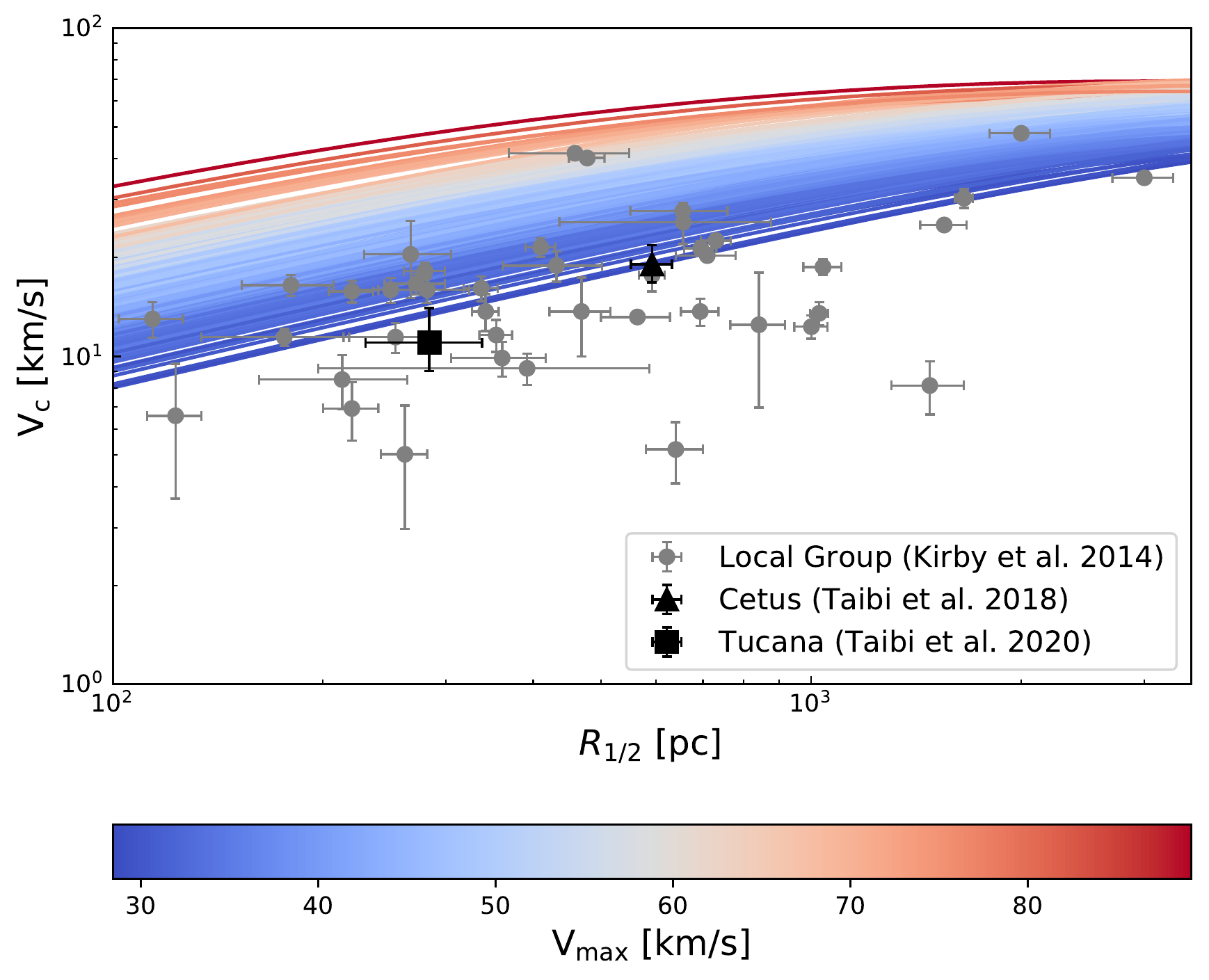}
		\caption{Circular velocity at the half light radius (under assumption of flat $\sigma$ profile, isotropy and sphericity) for Local Group galaxies from \protect{\citet{Kirby2014}} and the isolated dwarfs Cetus and Tucana, from \citet{Taibi2018} and this work. Circular velocity profiles for NFW halos between $8.5 \leq {\rm log} (M_{\rm vir}/M_\odot) \leq 11$ and following a mass concentration relation from \protect{\citet{Dutton14}} are shown color coded by $V_{\rm max}$. Local Group galaxies represented as gray filled circles.
		Cetus and Tucana are highlighted with a black triangle and square, respectively; they occupy locations comparable to other dwarfs of similar stellar mass (see also Fig.~\ref{fig:tbtf2}).
		}
		\label{fig:tbtf}
	\end{figure}
	
	We show in Figure \ref{fig:tbtf} the position of Tucana on the $V_{\rm circ} - r_{1/2}$ plane with respect to the Local Group compilation of \citet{Kirby2014} and the updated value for Cetus from \citet{Taibi2018}. NFW halos sampled from the halo mass concentration relation of \cite{Dutton14} are also shown (see Appendix \ref{sec:C} for sampling details) and color coded by $V_{\rm max}$. The updated velocity dispersion measurements for Tucana and Cetus place these galaxies in a locus occupied by comparable luminosity dwarf galaxies.
	
	These results would imply not only that Tucana does not reside in a very centrally dense halo as predicted by G19, but also that this galaxy it is not an exception to the too-big-to-fail problem (see e.g. \citealp{Kirby2014})
	
	Cetus and Tucana's isolation and well measured SFHs offer an intriguing leverage to potentially separate the multitude of solutions to the too-big-to-fail problem. 
	In particular, what can be expected from any environmental or SFH dependence on the galaxy density profile in baryonic feedback scenarios (see discussion in \citealt{Read19}), and how this could contrast with self-interacting dark matter solutions, which could predict more homogeneous behavior among all dwarfs. 
	Further assessing these scenarios in light of the results for the isolated dSphs, and their spatial stellar population distributions will be the focus of a follow-up work.

	\section{Chemical analysis} \label{sec:chemical}
	\subsection{Metallicity properties} \label{sec:metallicity}
	The analysis of the [Fe/H] values of the FORS2 combined dataset led to the following results: considering those values having $P_{\rm M}>0.95$, we obtained median [Fe/H]$=-1.58$~dex, $\sigma_{\rm MAD} = 0.47$~dex, $\sigma_{\rm intrinsic} = 0.39$~dex; while adding those stars with $0.05<P<0.95$ we got instead median [Fe/H]$=-1.61$~dex, $\sigma_{\rm MAD} = 0.48$~dex, $\sigma_{\rm intrinsic} = 0.39$~dex. 
	Therefore, Tucana is a metal-poor system with a significant spread in metallicity. 
	The median [Fe/H] value measured for the likely members is in very good agreement with the integrated quantity derived from Tucana's SFH: $\left \langle\textrm{[Fe/H]}\right \rangle = -1.52 \pm 0.07$~dex \citep{Monelli2010}. In addition, our average [Fe/H] value falls within the rms scatter of the stellar luminosity-metallicity relation for LG dwarf galaxies reported by \citet{Kirby2013}, while the intrinsic scatter agrees well with the values of other similarly luminous dwarf galaxies \citep{Leaman2013}.
	
	\begin{figure}
		\centering
		\includegraphics[width=\hsize]{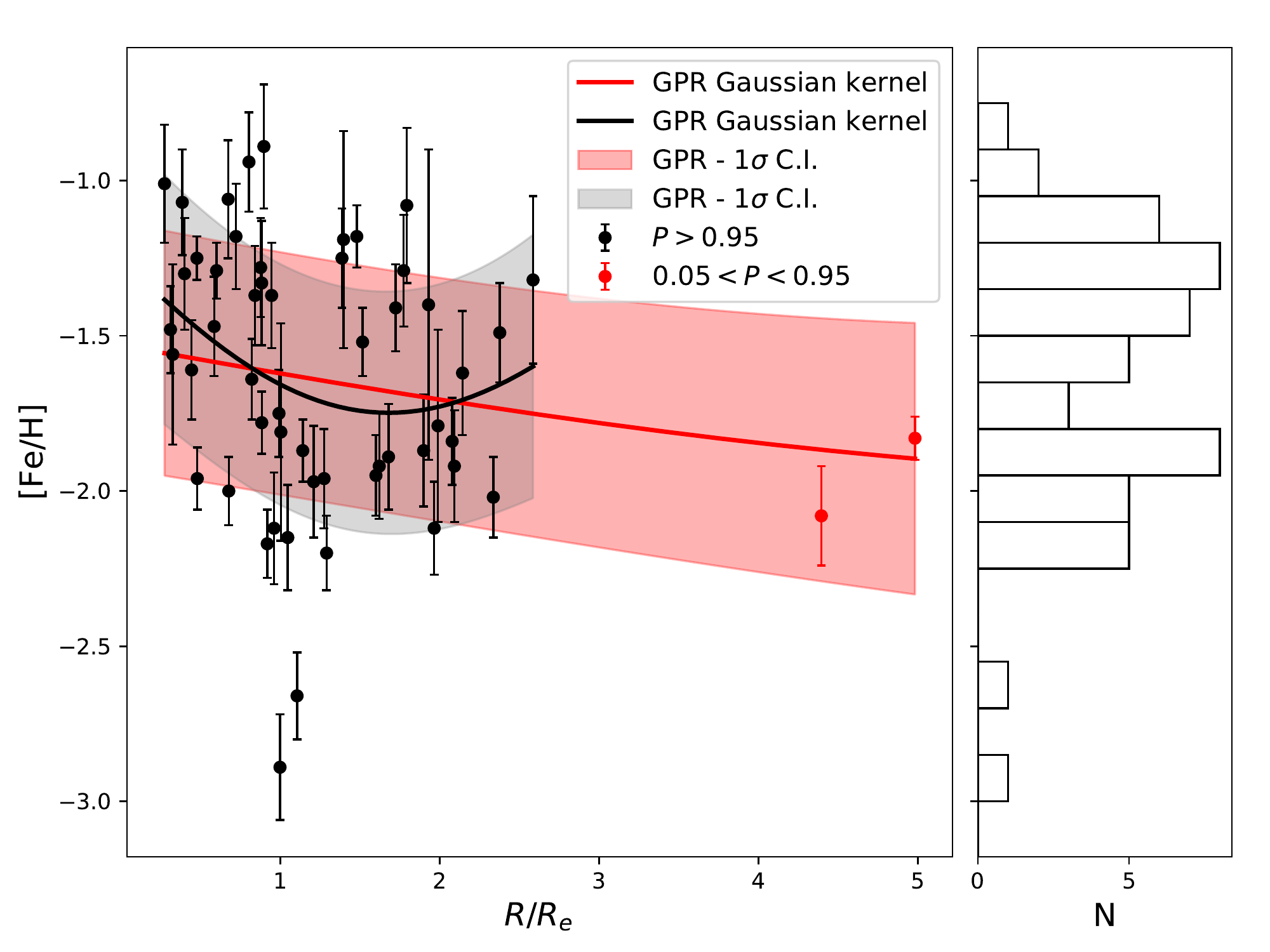}
		\caption{[Fe/H] as a function of the elliptical radius scaled with $R_e$ for Tucana's probable member stars from the FORS2 combined dataset. Black dots represent the targets with membership probabilities $P>0.95$, while the red dots those having $0.05<P<0.95$. The black solid line represents the result of a Gaussian process regression analysis using a Gaussian kernel and taking into account an intrinsic scatter; the gray band indicates the corresponding 1-$\sigma$ confidence interval. The red solid line and the red band indicate the same but using all targets with $P>0.05$. The histogram on the right side represents the metallicity distribution of the stars with $P>0.95$.
		}
		\label{Fig:met1.1}
	\end{figure}
	
	The distribution of the [Fe/H] values as a function of the elliptical radius is shown in Fig.~\ref{Fig:met1.1}. The significant scatter in metallicity toward the inner part of the galaxy is evident. In addition, there is a bi-modality in the metallicity histogram suggesting the presence of two sub-populations. 
	This can be related to the results obtained from deep photometric data \citep{Monelli2010}, where it has been shown that the splitting observed in the HB, in the RGB bump and in the properties of the RR-Lyrae stars, implies that Tucana has been able to produce a second generation of metal-rich stars thanks to the self-enrichment from the first stars in a very short period of star formation ($\sim1$~Gyr).
	In a more recent study, \citet{Savino2019} re-analyzed the photometric data from \citet{Monelli2010}, by studying the main-sequence turn-off and the HB of Tucana. They were able to obtain a SFH with a finer temporal resolution, showing that Tucana actually experienced two early phases of star formation (SF), followed by a third one  ending between 6 and 8 Gyr ago, with the two initial episodes being the most intensive.
	According to the age-metallicity relation recovered by \citet{Savino2019}, our [Fe/H] measurements can be related to the intermediate-old and intermediate-young age populations of the two last episodes of SF. This would explain the bi-modality we observed in the metallicity histogram, while our most metal-poor stars could be related to the oldest episode of SF. However, despite the relatively high intensity of this SF period, we found very few stars with [Fe/H]$<-2.25$~dex. 
	Probably, this is related to the fact that for lower metallicity stars, due to weaker lines, a higher S/N is needed in order to get similar accuracy in [Fe/H] measurements. Furthermore, they tend to be more extended which require an extra attention during the sampling phase.
	
	Observations have shown that in many LG dwarf galaxies the young and metal-rich stars are more spatially concentrated than the old and metal-poor ones which display a more extended spatial distribution. Their overall radial distribution produces a decreasing metallicity gradient -- e.g. Fornax \citep{Battaglia2006, Leaman2013}, Phoenix \citep{Kacharov2017} -- which can eventually reach a plateau on the outside -- e.g. Sculptor \citep{Tolstoy2004}, VV~124 \citep{Kirby2013b}, Cetus (T18). 
	These results have also been reproduced by simulations \citep[e.g.][]{Schroyen2013, Revaz+Jablonka2018}, which have shown that the shape of these gradients strongly depends on the combination of the stellar mass, SFH and dynamical history of the system under consideration, although their strength could be influenced by merger events \citep{Benitez-Llambay2016} or environmental effects such as tidal stripping \citep{Sales2010}. 
	
	Therefore, we investigated the presence of a metallicity gradient as a function of radius, first focusing on the stars with $P>0.95$, which extended up to $\sim3\,R_e$ (see Fig.~\ref{Fig:met1.1}).  
	Performing an error-weighted linear least-square fit to the data, we obtained the value $m = \frac{\mathrm{d} \rm{[Fe/H]}}{\mathrm{d} R} = -0.16 \pm 0.09$ dex\,arcmin$^{-1}$ ($=-0.6 \pm 0.4$ dex\,kpc$^{-1} =-0.13 \pm 0.07$ dex\,$R_e^{-1}$, using the values reported in Table~\ref{table:1} for the conversions). 
	We also performed a Gaussian process regression (GPR) analysis, where we used a Gaussian kernel together with a noise component to account for the intrinsic metallicity scatter. The GPR has the advantage of being a kernel-based non-parametric probabilistic method which allows to compute empirical confidence intervals. Since we are looking for a smooth function, it performs better than a least-square fit to find the general trend in the data. In our case we confirmed the decreasing trend, although the 1-$\sigma$ confidence limits resulted quite large due to the high intrinsic scatter of the data, making the presence of a metallicity gradient within $\sim3R_e$ dubious. 
	
	We further checked this result by performing a simple simulation. We assumed a double Gaussian metallicity distribution with parameters roughly fitting the observed one, but no spatial variation  ($\mu_{[Fe/H],1}=-2.0$~dex, $\sigma_{[Fe/H],1}=0.2$, $\mu_{[Fe/H],2}=-1.3$~dex, $\sigma_{[Fe/H],2}=0.2$ assuming the same fraction of stars in the two Gaussians). We then randomly extracted [Fe/H] values at the radial positions of our data. We further reshuffled the [Fe/H] values according to the observed errors and finally performed a linear least-square fit looking for a spatial metallicity gradient. We repeated this process 1000 times. 
	The obtained average gradient was compatible with zero, with the associated scatter large enough to include within 1-$\sigma$ the observed value of \textit{m}. Therefore, with the data at our hands, the observed gradient within $R<3\,R_e$ is not statistically significant.

	Adding the stars with $0.05<P<0.95$ would extend the spatial coverage up to $R\sim6R_e$, thanks to the two outermost targets, but would lead to an even milder gradient: $m = -0.07 \pm 0.04$ dex\,arcmin$^{-1}$ ($=-0.28 \pm 0.16$ dex\,kpc$^{-1} =-0.06 \pm 0.03$ dex\,$R_e^{-1}$), by performing a linear least-square fit. 	
	The presence of a metallicity gradient in Tucana is expected from studies of deep-photometric data \citep{Hidalgo2013,Savino2019}. However it is probable that we are mainly targeting stars belonging to the more recent episodes of SF, whose populations share similar spatial extensions (see e.g. Fig.11b in \citealp{Savino2019}).
	Therefore, the presence of a metallicity gradient in Tucana is very tentative and we would need a better sampling, in particular of the metal-poor component, around $3 \lesssim R/R_e \lesssim5$ to put our results on a firmer ground.
	
	We compared Tucana's metallicity gradient (or rather the lack of) with those of some MW satellites having similar luminosities ($L_V\lesssim 5 \time 10^5 L_\odot$) and short SFHs, i.e. Draco \citep{Aparicio2001}, Ursa Minor \citep{Carrera2002} and Sextans \citep{Bettinelli2018}. All of them have formed the majority of their stars more than 10 Gyr ago within a short period of SF, which in some cases may have lasted no more than 1 Gyr (i.e. Sextans). It has been shown that such short SFHs may lead to mild metallicity gradients in these systems \citep{Marcolini2008,Kirby2011,Revaz+Jablonka2018}. 
	Indeed, both Draco and Ursa Minor show mild gradients having values of $-0.05$ dex\,$R_e^{-1}$ and $-0.03$ dex\,$R_e^{-1}$, as reported by \citet{Schroyen2013} and \citet{Kirby2011}, respectively\footnote{We revised the Draco's gradient using the more recent and spatially extended dataset of \citet{Walker2015}: we performed a broad membership selection as in \citet{Walker2015} (see their Fig.~10) and then refined it by cross-correlating with the Gaia-DR2 catalog, selecting those targets co-moving with Draco. We found from the linear least-square fit a value of $-0.09 \pm 0.02$ dex\,$R_e^{-1}$, in fair agreement with the previous one.}.
	Sextans, on the other hand, seems to have a stronger gradient of $-0.24$ dex\,$R_e^{-1}$, as reported by \citet{Schroyen2013} using the \citet{Battaglia2011} spectroscopic dataset. However, this value is somewhat overestimated since recent studies of the structural properties of Sextans \citep{Roderick2016,Cicuendez2018} have shown that this system is less extended than what was previously reported in the literature. Using the half-light radius value from \citet{Cicuendez2018} we find a lower gradient of $-0.18$ dex\,$R_e^{-1}$, which is still far from the other dwarf's values and probably related to an early merger event that could have steepened its metallicity gradient (see \citealp{Cicuendez+Battaglia2018}, but also \citealp{Benitez-Llambay2016}).
	If the case of Tucana is similar to that of Draco and Ursa Minor, as it seems, we would expect it to host as much a mild metallicity gradient, but it would take a better sampling of the spatial extension of the metal-poor component in Tucana to confirm it. 
	
	\subsection{Searching for two chemo-kinematically distinct populations}\label{subsec:met-multipops}
	
	Although we have not spotted a clear metallicity gradient in Tucana, the bi-modality found in the metallicity distribution, may indicate the presence of two sub-populations which differ not only in their chemical properties but also in their kinematics. 
	Some of the dSphs satellites of the MW, such as Sculptor, Fornax, Carina and Sextans, show such features where the metal-rich (usually more spatially concentrated) sub-population has a colder kinematics than the metal-poor (and more extended) one \citep[e.g.][]{Tolstoy2004,Battaglia2006,Battaglia2008b,Battaglia2011,Koch2008,Amorisco+Evans2012}. 
	Determining the chemo-kinematic properties of dSphs is of great interest not only to better understand their evolutionary path, but also to get an insight into their dark matter properties \citep[e.g.][]{Battaglia2008b,Walker+Penarrubia2011,Amorisco+Evans2012b,Strigari2018}.
	
	In the case of Tucana, we have first analyzed the combined FORS2 dataset with the $P>0.95$ cut applied. We have taken the median [Fe/H] value of $-1.58$~dex to split our sample into a metal-rich (MR) and a metal-poor (MP) sub-samples. We then run our code to obtain the kinematic parameters of both samples (see Sect.\ref{sec:kinematics}). Using the dispersion-only model, we found $\sigma_{\rm v,MR}=6.0^{+1.9}_{-1.6}$ km\,s$^{-1}$ and $\sigma_{\rm v,MP}=9.0^{+2.8}_{-2.4}$ km\,s$^{-1}$, which are at 1-$\sigma$ from each other.
	Including instead the $0.05<P<0.95$ data, we obtained:
	$\sigma_{\rm v,MR}=5.9^{+1.9}_{-1.6}$ km\,s$^{-1}$ and $\sigma_{\rm v,MP}=10.9^{+2.6}_{-2.2}$ km\,s$^{-1}$, which are instead at $\sim$2-$\sigma$ from each other. 	
	
	Therefore, there is a weak evidence of two chemo-kinematically distinct sub-populations in Tucana. Additional data, in particular including external parts of the galaxy are necessary to reach firm conclusions both regarding the presence of a metallicity gradient as well as the possible distinct chemo-kinematic populations.
	
	\begin{table}
		\caption{Parameters from the Bayesian kinematic analysis of the metal-rich (MR) and metal-poor (MP) sub-samples defined in Sect.\ref{subsec:met-multipops}.}
		\label{table:MR+MP}
		\centering          
		\begin{tabular}{c  c c  c c}    
			\hline\hline
			& \multicolumn{2}{c}{$P>0.95$} & \multicolumn{2}{c}{$P>0.05$} \\ 
			Sample & N & $\sigma_{\rm v}$ & N & $\sigma_{\rm v}$ \\ 
			&  & (km\,s$^{-1}$) &  & (km\,s$^{-1}$) \\
			\hline     
			MR & 26 & 6.0$^{+1.9}_{-1.6}$ & 27 &  5.9$^{+1.9}_{-1.6}$ \\
			MP & 26 & 9.0$^{+2.8}_{-2.4}$ & 27 & 10.9$^{+2.6}_{-2.2}$ \\           
			\hline
		\end{tabular}
	\end{table}
	
	\begin{figure}
		\centering
		\includegraphics[width=\hsize]{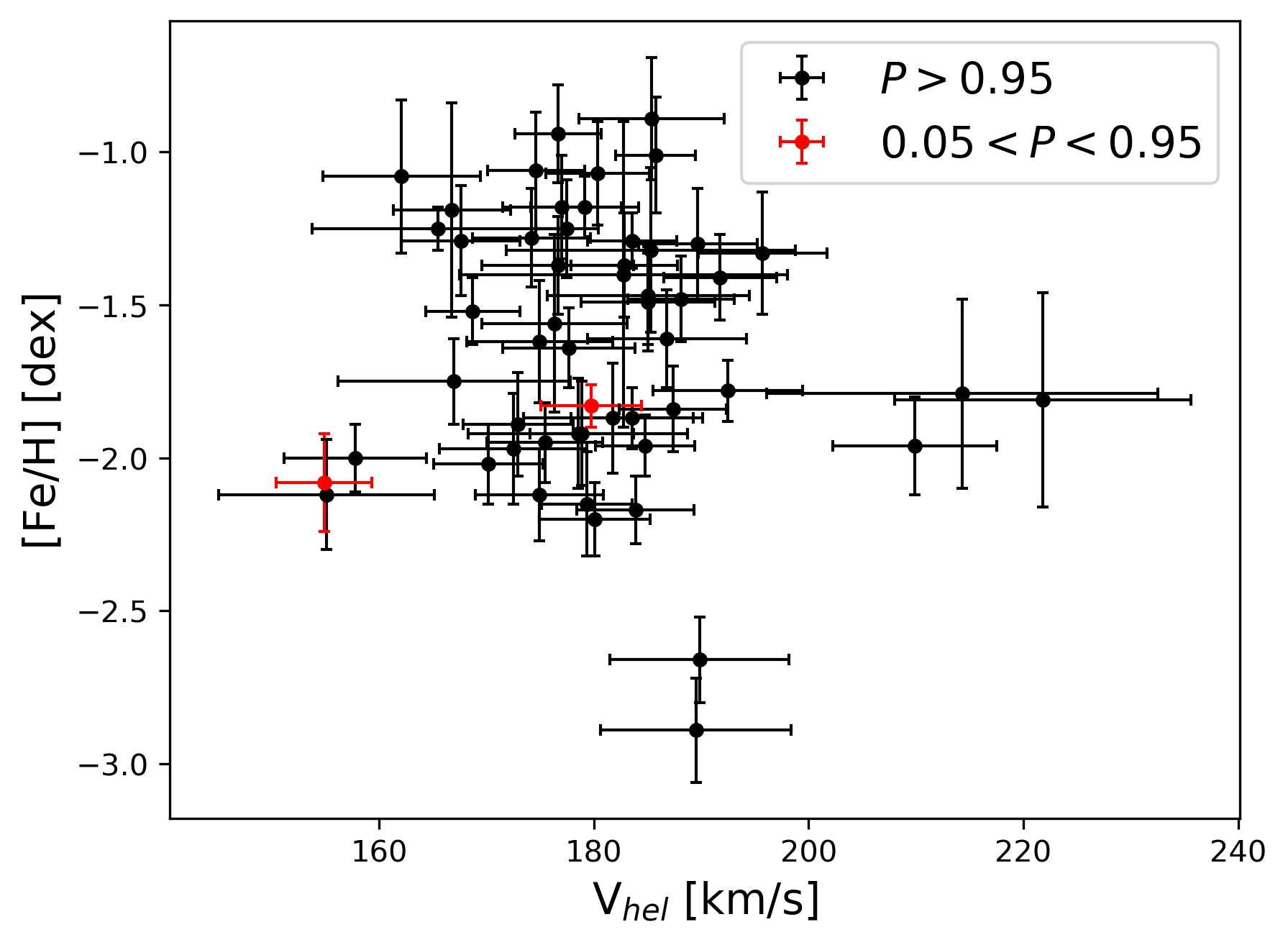}
		\caption{Metallicity versus heliocentric velocity for the observed target of the combined FORS2 dataset. Black dots represent the targets with $P>0.95$, while re dots those with $0.05<P<0.95$.
		}
		\label{Fig:met1.2}
	\end{figure}

	\section{Summary and Conclusions} \label{sec:summary}
	
	In this paper we present results on the internal kinematic and metallicity properties of the Tucana dwarf spheroidal galaxy, based on the analysis of multi-object spectroscopic samples of individual RGB stars.
	
	This analysis is based on a novel set of 50 individual objects collected with the VLT/FORS2 instrument in MXU mode in P91, complemented by a re-reduction and re-analysis of two datasets from the literature, i.e. the VLT/FORS2-MXU dataset presented in \citet[][F09]{Fraternali2009} and the VLT/FLAMES-GIRAFFE one by \citet[][G19]{Gregory2019}. 
	
	Applying a probabilistic membership approach, we find 39 effective members in our P91 sample, which doubles the number of Tucana's member stars found in F09 and G19. 
	
	A full re-reduction and analysis of the data presented in the literature was carried out because it became clear that the published catalogues could not be directly combined to the line-of-sight velocities we derived for the P91 sample: there are significant differences between the values of the systemic velocity reported in those studies with respect to that we derived from the P91 dataset ($\sim$195 km\,s$^{-1}$ for F09, $\sim$215 km\,s$^{-1}$ for G19, while $\sim$180 km\,s$^{-1}$ in our case); and the comparison of the individual line-of-sight velocities for the stars in common were supporting the presence of shifts with respect to F09, but were not sufficient to fully quantify whether that was the only source of difference, or was even more unfavorable for the comparison with G19. 
	
	Following our homogeneous data reduction, we find an excellent agreement between velocity measurements of the three datasets both for stars in common (Fig.~\ref{Fig:A.1} top row panels) as well as for systemic velocity, which is stable around 180 km\,s$^{-1}$ for the three datasets (see Fig.\ref{Fig:flames-hist}, right panel, and Table~\ref{table:multinest}).
	
	We proceeded to analyze the P91 dataset alone and also in combination with our treatment of the P69 and FLAMES data. 	
	The resulting values of the intrinsic l.o.s. velocity dispersion are consistently around 6 km\,s$^{-1}$ when considering the 3 combination of datasets (P91, P91$+$P69, all the three combined) and the highly probable members (probability of membership $P_{\rm M}>0.95$); when including lower probability members, the velocity dispersion increases, but it is unlikely that $\sigma_{\rm v}$ is larger than 10 km\,s$^{-1}$.
	
	Therefore, our analysis leads to the conclusion that the l.o.s. velocity dispersion of Tucana's stellar component is much lower than the values reported by F09 and G19 -- $\sigma_{\rm v,F09}=15.8^{+4.1}_{-3.1}$ km\,s$^{-1}$ and $\sigma_{\rm v,G19}=14.4^{+2.8}_{-2.3}$ km\,s$^{-1}$.
	
	Furthermore, we find no significant signs of internal rotation. Mock tests suggest that if Tucana would have had a maximum rotational velocity of $\sim$10-15 km\,s$^{-1}$ along the projected major axis (like what previously reported in literature) with the data at our hands we should have detected it with high significance. On the other hand, lower levels of rotation are not completely ruled out, but a larger sample would be needed to quantify their presence. Nevertheless it seems improbable that Tucana is a fast rotator ($v_{\rm rot}/\sigma_{\rm v}\gtrsim1$).
	
	Assuming for our data an average $\sigma_{\rm v} = 6.2_{-1.3}^{+1.6}$ km\,s$^{-1}$, we obtain a dynamical mass within the half-light radius of $M_{1/2}=0.7_{-0.3}^{+0.4}\times 10^7 M_\odot$. This translates into a circular velocity at the half-light radius of $V_{\rm circ}(r_{1/2}) = 11_{-2}^{+3}$ km\,s$^{-1}$ which implies that, if Tucana inhabits a NFW dark matter halo, it should have a similar density as those of other MW dSphs \citep{Boylan-Kolchin2012}. Therefore, Tucana is not an exception to the too-big-to-fail problem and not "a massive failure", as it had gained fame of.
	
	The analysis of Tucana's chemical properties has been carried out only on the combined P91 and P69 FORS2 data, due to their higher S/N. 
	We establish that the galaxy is mainly metal-poor with a significant scatter in metallicity (having a median [Fe/H] $=-1.58$ dex, $\sigma_{{\rm MAD}}= 0.47$ dex and $\sigma_{{\rm intrinsic}}= 0.39$ dex when considering only highly likely members, and median [Fe/H] $=-1.61$ dex, $\sigma_{{\rm MAD}}= 0.48$ dex and $\sigma_{{\rm intrinsic}}= 0.39$ dex when including the less likely members). The derived values agree very well with SFH studies \citep{Monelli2010, Savino2019}.  In addition the average [Fe/H] falls between the rms scatter of the stellar luminosity-metallicity relation for LG dwarf galaxies \citep{Kirby2013}.
	
	Looking at the distribution of the [Fe/H] values as a function of radius, we find a mild metallicity gradient. However, the size and spatial distribution of the current datasets do not lead to a statistically significant detection. The presence of a gradient in Tucana would be expected from the age gradients inferred from deep photometric studies \citep{Hidalgo2013, Savino2019}, but also from both observations and simulations of similarly luminous dwarfs \citep[see e.g.][]{Leaman2013, Schroyen2013, Revaz+Jablonka2018}, which indeed host mild metallicity gradients.
	Therefore, the presence of an underling gradient in Tucana is not excluded, but it would need a better sampling of the metal-poor component in Tucana (particularly at $R\gtrsim3R_e$) to confirm it.
	
	Finally, we find a hint of the presence of multiple stellar populations having distinct chemo-kinematical properties, although also in this case the addition of new data would help to put this result on a firmer ground. 
	
	\begin{acknowledgements}
		We would like to thank the anonymous referee for the helpful and constructive comments.		
		
		We thank Giacomo Beccari and Michele Bellazzini for sharing with us their VLT/VIMOS photometric data. We also thank F. Fraternali, E. Tolstoy, A. Gregory and M. Collins for useful discussions in the different phases of this project.
		
		S.T. thankfully acknowledges ESO for a one month visit funded by an SSDF grant.
		G.B. and S.T. acknowledge financial support through the grants (AEI/FEDER, UE) AYA2017-89076-P, AYA2014-56795-P, as well as by the Ministerio de Ciencia, Innovaci\'on y Universidades (MCIU), through the State Budget and by the Consejer\'\i a de Econom\'\i a, Industria, Comercio y Conocimiento of the Canary Islands Autonomous Community, through the Regional Budget. S.T. acknowledges an FPI fellowship BES-2015-074765, while G.B. acknowledges financial support through the grant RYC-2012-11537.
		
		This research has made use of NASA's Astrophysics Data System and extensive use of IRAF, Python, Numpy, Scipy and Astropy ecosystems.
	\end{acknowledgements}
	
	%
	%
	
	\bibliographystyle{bibtex/aa} 
	\bibliography{bibtex/tucana.bib} 
	
	\longtab{
		\begin{longtable}{rrlccccclll}
			\caption{\label{table:P91}
				Properties of the observed P91-FORS2 dataset in the Tucana dSph. Column (1) field Tuc0 and Tuc1; (2) slit aperture: numbers < 30 indicate observed targets from chip-1, otherwise from chip-2 -- numbers counted from bottom to top of the CCD; (3) RA-Dec coordinated in J2000; (4) V band magnitude with error from VLT/FORS2 photometric catalog (5) I band magnitude with error from same catalog (6) l.o.s. heliocentric velocity with error; (6) metallicity with error; (8) S/N ratio in pxl$^{-1}$ (the conversion factor to \AA$^{-1}$ is 1.09);(9) probability of membership -- the three columns indicate the probabilities obtained using the P91 dataset alone ($P_1$), combining with the P69 ($P_2$) and further adding the FLAMES data($P_3$). Four stars had repeated measurements: targets 31, 32, 33 and 35 from Tuc0 field corresponding to targets 31, 32, 33 and 36 from Tuc1 field, respectively; in this table we report the single measurements as well as the averaged values used during the analysis process.}\\
			\hline\hline
			Field & Slit &  Ra, Dec (J2000)& $V\pm\delta V$ & $I\pm\delta I$ & $v_{hel} \pm \delta v_{hel}$ & ${\rm [Fe/H]}\pm \delta_{\rm [Fe/H]}$ & SNR &  \multicolumn{3}{c}{P}  \\
			&  &  [deg] & &        & [km/s]  & [dex] & [pxl$^{-1}$] & P$_{1}$ & P$_{2}$ & P$_{3}$\\
			\hline
			\endfirsthead
			\caption{continued.}\\
			\hline\hline
			Field & Slit &  Ra, Dec (J2000) & $V\pm\delta V$ & $I\pm\delta I$ & $v_{hel} \pm \delta v_{hel}$ & ${\rm [Fe/H]}\pm \delta_{\rm [Fe/H]}$  & SNR &  \multicolumn{3}{c}{P}  \\
			&  & [deg]  & &        & [km/s]  & [dex] & [pxl$^{-1}$] & P$_{1}$ & P$_{2}$ & P$_{3}$  \\
			\hline
			\endhead
			\hline
			\endfoot
			0  &   1  &  340.4913, -64.41896 & 22.79 $\pm$ 0.02 & 21.44 $\pm$ 0.04 & 176.6 $\pm$  7.1 & $-$1.37 $\pm$ 0.16 & 22.5 & 0.99 & 0.99 & 0.99 \\
			0  &   2  &  340.4834, -64.41816 & 22.75 $\pm$ 0.03 & 21.53 $\pm$ 0.04 & 157.8 $\pm$  6.6 & $-$2.00 $\pm$ 0.11 & 21.4 & 0.99 & 0.99 & 0.99 \\
			0  &   3  &  340.4762, -64.41057 & 22.33 $\pm$ 0.02 & 21.19 $\pm$ 0.04 & 189.8 $\pm$  8.3 & $-$2.66 $\pm$ 0.14 & 29.4 & 0.99 & 0.99 & 0.99 \\
			0  &   4  &  340.4645, -64.42837 & 22.69 $\pm$ 0.02 & 21.33 $\pm$ 0.04 & 183.9 $\pm$  5.5 & $-$2.17 $\pm$ 0.11 & 24.5 & 0.99 & 0.99 & 0.99 \\
			0  &   5  &  340.4565, -64.42895 & 22.88 $\pm$ 0.02 & 21.56 $\pm$ 0.03 & 167.0 $\pm$ 10.8 & $-$1.75 $\pm$ 0.14 & 19.4 & 0.99 & 0.99 & 0.99 \\
			0  &   6  &  340.4511, -64.42363 & 22.19 $\pm$ 0.02 & 20.55 $\pm$ 0.03 & 165.5 $\pm$ 11.7 & $-$1.25 $\pm$ 0.07 & 12.2 & 0.99 & 0.99 & 0.99 \\
			0  &   7  &  340.4421, -64.42360 & 22.29 $\pm$ 0.02 & 20.87 $\pm$ 0.03 & 183.6 $\pm$  4.1 & $-$1.29 $\pm$ 0.09 & 38.3 & 0.99 & 0.99 & 0.99 \\
			0  &   8  &  340.4334, -64.42592 & 22.81 $\pm$ 0.03 & 21.57 $\pm$ 0.03 & 182.8 $\pm$  4.9 & $-$1.37 $\pm$ 0.17 & 16.2 & 0.99 & 0.99 & 0.99 \\
			0  &   9  &  340.4283, -64.41176 & 22.72 $\pm$ 0.02 & 21.46 $\pm$ 0.04 & 155.1 $\pm$ 10.0 & $-$2.12 $\pm$ 0.18 & 21.3 & 0.99 & 0.99 & 0.99 \\
			0  &  10  &  340.4199, -64.41700 & 22.83 $\pm$ 0.02 & 21.49 $\pm$ 0.04 & 174.2 $\pm$  5.5 & $-$1.28 $\pm$ 0.16 & 23.1 & 0.99 & 0.99 & 0.99 \\
			0  &  11  &  340.4090, -64.41599 & 22.34 $\pm$ 0.02 & 20.91 $\pm$ 0.03 & 183.6 $\pm$  5.7 & $-$1.87 $\pm$ 0.10 & 36.3 & 0.99 & 0.99 & 0.99 \\
			0  &  12  &  340.3984, -64.41116 & 22.44 $\pm$ 0.02 & 20.96 $\pm$ 0.04 & 168.7 $\pm$  4.4 & $-$1.52 $\pm$ 0.11 & 29.0 & 0.99 & 0.99 & 0.99 \\
			0  &  13  &  340.3879, -64.41004 & 22.89 $\pm$ 0.02 & 21.55 $\pm$ 0.04 & 167.6 $\pm$  5.5 & $-$1.29 $\pm$ 0.18 & 20.7 & 0.99 & 0.99 & 0.99 \\
			0  &  14  &  340.3760, -64.42418 & 22.67 $\pm$ 0.02 & 21.36 $\pm$ 0.04 & 187.4 $\pm$  5.0 & $-$1.84 $\pm$ 0.14 & 26.6 & 0.99 & 0.99 & 0.99 \\
			0  &  15  &  340.3693, -64.41906 & 22.89 $\pm$ 0.02 & 21.82 $\pm$ 0.04 & 178.5 $\pm$ 10.2 & $-$1.92 $\pm$ 0.18 & 18.8 & 0.99 & 0.99 & 0.99 \\
			0  &  16  &  340.3576, -64.41379 & 22.39 $\pm$ 0.02 & 20.17 $\pm$ 0.03 &  43.3 $\pm$  5.8 & $-$2.05 $\pm$ 0.05 & 54.0 & Ph   &      & \\   
			& rep. &  340.6158, -64.40893 & 22.96 $\pm$ 0.02 & 21.36 $\pm$ 0.04 & 146.1 $\pm$  7.0 & $-$1.88 $\pm$ 0.12 & 27.0 & 0.02 & 0.02 & 0.03 \\
			0  &  31  &   &    &    &  147.5 $\pm$ 13.6  &  -1.81 $\pm$ 0.18  &  25.2  &   &      &  \\
			1  &  31  &   &    &    &  145.6 $\pm$  8.3  &  -1.94 $\pm$ 0.17  &  28.9  &   &      &  \\
			& rep. &  340.6056, -64.4071  & 21.65 $\pm$ 0.02 & 21.29 $\pm$ 0.03 &    /             &    /             & 24.3 & Ph   &      & \\
			0  &  32  &   &    &    &   /             &    /             &  22.7  &   &      &  \\
			1  &  32  &   &    &    &   /             &    /             &  25.9  &   &      &  \\            
			& rep. &  340.5832, -64.43912 & 22.95 $\pm$ 0.02 & 21.36 $\pm$ 0.03 & 225.8 $\pm$  4.4 & $-$1.67 $\pm$ 0.09 & 26.5 & 0.00 & 0.00 & 0.00 \\
			0  &  33  &   &    &    &  221.8 $\pm$  7.2  &  -1.75 $\pm$ 0.12  &  24.3  &   \\
			1  &  33  &   &    &    &  228.3 $\pm$  5.6  &  -1.55 $\pm$ 0.15  &  28.6  &   \\
			0  &  34  &  340.5744, -64.41475 & 23.63 $\pm$ 0.03 & 22.46 $\pm$ 0.05 &    /             &    /             &  7.7 & Ph   &      & \\
			& rep. &  340.5692, -64.38342 & 22.39 $\pm$ 0.02 & 20.62 $\pm$ 0.03 & 179.8 $\pm$  4.7 & $-$1.83 $\pm$ 0.07 & 35.0 & 0.89 & 0.88 & 0.88 \\
			0  &  35  &   &    &    &  181.0 $\pm$  6.4  &  -1.88 $\pm$ 0.11  &  32.5  &   &      &  \\
			1  &  36  &   &    &    &  178.3 $\pm$  6.9  &  -1.79 $\pm$ 0.09  &  37.5  &   &      &  \\
			0  &  36  &  340.5508, -64.43098 & 22.78 $\pm$ 0.02 & 21.50 $\pm$ 0.04 & 185.0 $\pm$  6.2 & $-$1.49 $\pm$ 0.16 & 20.2 & 0.99 & 0.99 & 0.99 \\
			0  &  37  &  340.5418, -64.41572 & 23.23 $\pm$ 0.02 & 22.03 $\pm$ 0.05 & 175.0 $\pm$  6.8 & $-$1.62 $\pm$ 0.20 & 14.7 & 0.99 & 0.99 & 0.99 \\
			0  &  38  &  340.5302, -64.43026 & 23.65 $\pm$ 0.03 & 22.63 $\pm$ 0.08 & 182.8 $\pm$ 15.3 & $-$1.40 $\pm$ 0.50 &  8.2 & 0.99 & 0.99 & 0.99 \\
			0  &  39  &  340.5278, -64.40337 & 23.15 $\pm$ 0.02 & 21.92 $\pm$ 0.04 & 185.3 $\pm$ 13.4 & $-$1.32 $\pm$ 0.27 & 16.1 & 0.99 & 0.99 & 0.99 \\
			0  &  40  &  340.5186, -64.41192 & 22.90 $\pm$ 0.03 & 21.58 $\pm$ 0.05 & 162.1 $\pm$  7.3 & $-$1.08 $\pm$ 0.25 & 21.6 & 0.99 & 0.99 & 0.99 \\
			0  &  41  &  340.5072, -64.42974 & 22.39 $\pm$ 0.02 & 20.84 $\pm$ 0.03 & 179.2 $\pm$  5.0 & $-$1.18 $\pm$ 0.10 & 30.8 & 0.99 & 0.99 & 0.99 \\
			1  &   1  &  340.5224, -64.43385 & 23.46 $\pm$ 0.03 & 22.25 $\pm$ 0.07 & 214.3 $\pm$ 18.2 & $-$1.79 $\pm$ 0.31 &  9.6 & 0.98 & 0.99 & 0.99 \\
			1  &   2  &  340.5116, -64.44016 & 22.67 $\pm$ 0.02 & 21.36 $\pm$ 0.05 & 170.2 $\pm$  5.1 & $-$2.02 $\pm$ 0.13 & 23.7 & 0.99 & 0.99 & 0.99 \\
			1  &   3  &  340.5134, -64.41830 & 22.78 $\pm$ 0.02 & 21.44 $\pm$ 0.05 & 177.5 $\pm$  2.9 & $-$1.25 $\pm$ 0.16 & 26.6 & 0.99 & 0.99 & 0.99 \\
			1  &   4  &  340.4990, -64.42399 & 22.63 $\pm$ 0.02 & 21.33 $\pm$ 0.03 & 179.4 $\pm$  4.2 & $-$2.15 $\pm$ 0.17 & 26.6 & 0.99 & 0.99 & 0.99 \\
			1  &   5  &  340.4922, -64.42342 & 22.63 $\pm$ 0.02 & 21.41 $\pm$ 0.04 & 192.5 $\pm$  6.9 & $-$1.78 $\pm$ 0.10 & 26.6 & 0.99 & 0.99 & 0.99 \\
			1  &   6  &  340.4890, -64.41446 & 23.31 $\pm$ 0.02 & 22.12 $\pm$ 0.06 & 221.8 $\pm$ 13.8 & $-$1.81 $\pm$ 0.35 & 13.2 & 0.98 & 0.98 & 0.98 \\
			1  &   7  &  340.4818, -64.41523 & 22.75 $\pm$ 0.02 & 21.33 $\pm$ 0.04 & 176.7 $\pm$  4.0 & $-$0.94 $\pm$ 0.16 & 27.9 & 0.99 & 0.99 & 0.99 \\
			1  &   8  &  340.4706, -64.42798 & 22.86 $\pm$ 0.02 & 21.54 $\pm$ 0.05 & 185.4 $\pm$  6.8 & $-$0.89 $\pm$ 0.20 & 22.0 & 0.99 & 0.99 & 0.99 \\
			1  &   9  &  340.4658, -64.42591 & 22.78 $\pm$ 0.02 & 21.49 $\pm$ 0.03 & 174.6 $\pm$  4.5 & $-$1.06 $\pm$ 0.19 & 20.3 & 0.99 & 0.99 & 0.99 \\
			1  &  10  &  340.4619, -64.41730 & 22.60 $\pm$ 0.02 & 21.29 $\pm$ 0.03 & 185.8 $\pm$  3.7 & $-$1.01 $\pm$ 0.19 & 25.6 & 0.99 & 0.99 & 0.99 \\
			1  &  11  &  340.4553, -64.41642 & 22.42 $\pm$ 0.02 & 21.12 $\pm$ 0.03 & 188.1 $\pm$  5.0 & $-$1.48 $\pm$ 0.14 & 33.6 & 0.99 & 0.99 & 0.99 \\
			1  &  12  &  340.4446, -64.42166 & 22.73 $\pm$ 0.03 & 21.47 $\pm$ 0.03 & 189.7 $\pm$  5.5 & $-$1.30 $\pm$ 0.18 & 22.3 & 0.99 & 0.99 & 0.99 \\
			1  &  13  &  340.4351, -64.42136 & 22.59 $\pm$ 0.02 & 21.24 $\pm$ 0.03 & 185.1 $\pm$  9.4 & $-$1.47 $\pm$ 0.16 & 26.2 & 0.99 & 0.99 & 0.99 \\
			1  &  14  &  340.4321, -64.41439 & 22.61 $\pm$ 0.02 & 21.15 $\pm$ 0.03 & 177.0 $\pm$  5.5 & $-$1.18 $\pm$ 0.17 & 27.7 & 0.99 & 0.99 & 0.99 \\
			1  &  15  &  340.4200, -64.42276 & 22.53 $\pm$ 0.02 & 21.24 $\pm$ 0.04 & 189.5 $\pm$  8.9 & $-$2.89 $\pm$ 0.17 & 24.7 & 0.99 & 0.99 & 0.99 \\
			1  &  16  &  340.4267, -64.39091 & 22.43 $\pm$ 0.02 & 20.59 $\pm$ 0.03 &  -8.5 $\pm$  4.1 & $-$1.88 $\pm$ 0.10 & 48.0 & 0.00 & 0.00 & 0.00 \\
			1  &  17  &  340.4098, -64.41052 & 23.05 $\pm$ 0.02 & 21.79 $\pm$ 0.05 &    /             &    /             & 16.1 & Ph   &      & \\ 
			1  &  18  &  340.4026, -64.40694 & 22.67 $\pm$ 0.02 & 21.35 $\pm$ 0.04 & 172.9 $\pm$  5.1 & $-$1.89 $\pm$ 0.17 & 25.9 & 0.99 & 0.99 & 0.99 \\
			1  &  19  &  340.3919, -64.41256 & 22.70 $\pm$ 0.02 & 21.36 $\pm$ 0.04 & 175.5 $\pm$  5.4 & $-$1.95 $\pm$ 0.13 & 23.7 & 0.99 & 0.99 & 0.99 \\
			1  &  20  &  340.3787, -64.43124 & 22.22 $\pm$ 0.02 & 22.19 $\pm$ 0.05 &    /             &    /             &  7.9 & Ph   &      & \\
			1  &  34  &  340.5865, -64.38883 & 20.16 $\pm$ 0.01 & 19.47 $\pm$ 0.02 &    /             &    /             & 49.0 & Ph   &      & \\  
			1  &  35  &  340.5724, -64.39045 & 23.15 $\pm$ 0.03 & 20.41 $\pm$ 0.04 &    /             &    /             & 27.3 & Ph   &      & \\ 
			1  &  37  &  340.5571, -64.38788 & 22.93 $\pm$ 0.02 & 21.28 $\pm$ 0.05 & 154.9 $\pm$  4.5 & $-$2.08 $\pm$ 0.16 & 29.1 & 0.06 & 0.07 & 0.08 \\ 
			\hline    
		\end{longtable}
		\tablefoot{In the P columns, stars marked as "Ph" are non-members excluded according to their magnitudes and colors or because without reliable measurements (see Sect.~\ref{sec:membership} for full description)}
	}
	
	\longtab{
		\begin{longtable}{llccccclll}
			\caption{\label{table:P69}
				Properties of the observed P69-FORS2 dataset in the Tucana dSph. Column (1) slit aperture: numbers < 30 indicate observed targets from chip-1, otherwise from chip-2 -- numbers counted from bottom to top of the CCD; (2) RA-Dec coordinated in J2000; (3) V band magnitude with error from VLT/FORS2 photometric catalog (4) I band magnitude with error from same catalog (5) l.o.s. heliocentric velocity with error; (6) metallicity with error; (7) S/N ratio in pxl$^{-1}$ (the conversion factor to \AA$^{-1}$ is 1.08); (8) probability of membership according to the best-fitting kinematic model -- the three columns indicate the probabilities obtained using the P69 dataset alone ($P_1$), combining with the P91 ($P_2$) and further adding the FLAMES data($P_3$).}\\
			\hline\hline
			Slit &  Ra, Dec (J2000)& $V\pm\delta V$ & $I\pm\delta I$ & $v_{hel} \pm \delta v_{hel}$ & ${\rm [Fe/H]}\pm \delta_{\rm [Fe/H]}$ & SNR &  \multicolumn{3}{c}{P}  \\
			 &  [deg] & &        & [km/s]  & [dex] & [pxl$^{-1}$] & P$_{1}$ & P$_{2}$ & P$_{3}$\\
			\hline
			\endfirsthead
			\caption{continued.}\\
			\hline\hline
			Slit &  Ra, Dec (J2000) & $V\pm\delta V$ & $I\pm\delta I$ & $v_{hel} \pm \delta v_{hel}$ & ${\rm [Fe/H]}\pm \delta_{\rm [Fe/H]}$  & SNR &  \multicolumn{3}{c}{P}  \\
			 & [deg]  & &        & [km/s]  & [dex] & [pxl$^{-1}$] & P$_{1}$ & P$_{2}$ & P$_{3}$  \\
			\hline
			\endhead
			\hline
			\endfoot
			  1    & 340.4472, -64.42209 & 22.44 $\pm$ 0.02 & 20.97 $\pm$ 0.02 & 180.4 $\pm$ 4.8  & -1.07 $\pm$ 0.17 & 20.4 & 0.99 & 0.99 & 0.99 \\ 
			  2    & 340.4532, -64.42234 & 22.48 $\pm$ 0.02 & 21.26 $\pm$ 0.02 & 176.3 $\pm$ 6.8  & -1.56 $\pm$ 0.29 & 15.0 & 0.99 & 0.99 & 0.99 \\ 
			  3    & 340.4586, -64.39465 & 22.22 $\pm$ 0.02 & 20.96 $\pm$ 0.02 & 144.2 $\pm$ 5.2  & -1.52 $\pm$ 0.22 & 19.8 & 0.17 & 0.01 & 0.02 \\ 
			  4    & 340.4647, -64.4236  & 22.64 $\pm$ 0.01 & 21.22 $\pm$ 0.01 & 186.8 $\pm$ 7.4  & -1.61 $\pm$ 0.16 & 19.4 & 0.99 & 0.99 & 0.99 \\ 
			  5\tablefootmark{a}   & 340.4698, -64.43846 & 23.26 $\pm$ 0.03 & 21.50 $\pm$ 0.05 & 130.4 $\pm$ 10.1 & -1.37 $\pm$ 0.20 & 11.9 & 0.13 & 0.04 & 0.04 \\ 
			  6    & 340.4748, -64.41809 & 22.39 $\pm$ 0.01 & 20.87 $\pm$ 0.01 & 184.8 $\pm$ 4.6  & -1.96 $\pm$ 0.10 & 29.8 & 0.99 & 0.99 & 0.99 \\ 
			  7    & 340.4799, -64.40572 & 22.28 $\pm$ 0.02 & 20.85 $\pm$ 0.02 & 178.9 $\pm$ 4.8  & -1.92 $\pm$ 0.17 & 24.5 & 0.99 & 0.99 & 0.99 \\  
			  8    & 340.4881, -64.41732 & 22.73 $\pm$ 0.03 & 21.19 $\pm$ 0.03 & 177.7 $\pm$ 6.2  & -1.64 $\pm$ 0.13 & 18.9 & 0.99 & 0.99 & 0.99 \\   
			  9\tablefootmark{b}   & 340.4920, -64.42341 & 22.61 $\pm$ 0.02 & 21.35 $\pm$ 0.02 & 232.8 $\pm$ 8.0  & -1.28 $\pm$ 0.22 & 15.4 & Ph   &  &  \\ 
			  10   & 340.4963, -64.41156 & 22.45 $\pm$ 0.02 & 21.01 $\pm$ 0.02 & 225.0 $\pm$ 7.2  & -2.00 $\pm$ 0.18 & 19.8 & Ph   &  &  \\ 
			  11   & 340.5012, -64.41237 & 22.73 $\pm$ 0.01 & 21.49 $\pm$ 0.01 & 166.8 $\pm$ 5.4  & -1.19 $\pm$ 0.35 & 13.0 & 0.99 & 0.99 & 0.99 \\ 
			  12   & 340.5069, -64.40718 & 22.28 $\pm$ 0.02 & 20.86 $\pm$ 0.02 & 181.8 $\pm$ 8.3  & -1.87 $\pm$ 0.18 & 21.8 & 0.99 & 0.99 & 0.99 \\ 
			  13   & 340.5105, -64.42116 & 22.31 $\pm$ 0.02 & 20.83 $\pm$ 0.02 & 209.9 $\pm$ 7.6  & -1.96 $\pm$ 0.16 & 18.3 & 0.99 & 0.96 & 0.96 \\ 
			  14   & 340.5140, -64.41146 & 22.49 $\pm$ 0.03 & 20.96 $\pm$ 0.02 & 191.7 $\pm$ 5.3  & -1.41 $\pm$ 0.14 & 23.9 & 0.99 & 0.99 & 0.99 \\ 
			  23\tablefootmark{b}  & 340.5695, -64.3833  & 22.41 $\pm$ 0.03 & 20.65 $\pm$ 0.04 & 179.5 $\pm$ 5.7  & -1.95 $\pm$ 0.16 & 25.2 & 0.79 &  &  \\ 
			  24   & 340.5742, -64.37577 & 0.0   $\pm$ 0.0  & 0.0   $\pm$ 0.0  & 4.2 $\pm$ 4.8  & /                & 32.8 & 0.00 & 0.00 & 0.00 \\ 
			  25\tablefootmark{a,b} & 340.5836, -64.43902 & 22.95 $\pm$ 0.02 & 21.46 $\pm$ 0.03 & 236.4 $\pm$ 14.1 & -2.01 $\pm$ 0.26 & 17.4 & 0.63 &  &  \\ 
			  31   & 340.3294, -64.4244  & 21.78 $\pm$ 0.02 & 20.59 $\pm$ 0.02 & 208.6 $\pm$ 5.2  & /                & 23.5 & Ph   &  &  \\  
			  33\tablefootmark{a}  & 340.3435, -64.39041 & 22.61 $\pm$ 0.02 & 21.33 $\pm$ 0.02 & 110.3 $\pm$ 14.9 & -2.42 $\pm$ 0.28 & 14.6 & 0.00 & 0.00 & 0.00 \\   
			  35   & 340.3536, -64.41425 & 22.59 $\pm$ 0.02 & 21.0  $\pm$ 0.02 & 76.6 $\pm$ 7.3  & -1.77 $\pm$ 0.15 & 21.4 & 0.00 & 0.00 & 0.00 \\  
			  39\tablefootmark{a,b} & 340.3762, -64.4242  & 22.69 $\pm$ 0.02 & 21.38 $\pm$ 0.02 & 197.3 $\pm$ 8.3  & -2.11 $\pm$ 0.21 & 12.9 & 0.99 &  &  \\ 
			  40   & 340.3820, -64.45349 & 0.0   $\pm$ 0.0  & 0.0   $\pm$ 0.0  & 25.6 $\pm$ 11.1 & /                & 27.0 & 0.00 & 0.00 & 0.00 \\ 
			  43\tablefootmark{b}  & 340.3973, -64.41121 & 22.45 $\pm$ 0.02 & 20.98 $\pm$ 0.02 & 207.8 $\pm$ 7.3  & -1.14 $\pm$ 0.17 & 21.3 & Ph   &  &  \\ 
			  44   & 340.4024, -64.41617 & 22.27 $\pm$ 0.02 & 20.77 $\pm$ 0.02 & 180.1 $\pm$ 5.2  & -2.20 $\pm$ 0.12 & 25.7 & 0.99 & 0.99 & 0.99 \\ 
			  45   & 340.4100, -64.40206 & 22.41 $\pm$ 0.02 & 21.03 $\pm$ 0.02 & 174.9 $\pm$ 5.9  & -2.12 $\pm$ 0.15 & 21.3 & 0.99 & 0.99 & 0.99 \\ 
			  46\tablefootmark{a}  & 340.4163, -64.40805 & 21.09 $\pm$ 0.01 & 20.36 $\pm$ 0.01 & -160.6 $\pm$ 4.8  & /                & 31.7 & Ph   &  &  \\   
			  47   & 340.4206, -64.41588 & 22.68 $\pm$ 0.02 & 21.28 $\pm$ 0.02 & 195.7 $\pm$ 6.0  & -1.33 $\pm$ 0.20 & 14.7 & 0.99 & 0.99 & 0.99 \\ 
			  49   & 340.4348, -64.40812 & 22.37 $\pm$ 0.03 & 21.01 $\pm$ 0.02 & 172.5 $\pm$ 6.8  & -1.97 $\pm$ 0.18 & 24.3 & 0.99 & 0.99 & 0.99 \\ 
			\hline    
		\end{longtable}
		\tablefoot{In the P columns, stars marked as "Ph" are non-members excluded according to their magnitudes and colors or because their measurements resulted to be not reliable (see Sect.~\ref{sec:membership} and Appendix~\ref{sec:A} for full description). 
		\tablefoottext{a}{New with respect to \citet{Fraternali2009}}; \tablefoottext{b}{In common with the P91-FORS2 dataset}.}
	}
	
	\longtab{
		\begin{longtable}{l l c c c c c l l}
			\caption{\label{table:FLAMES}
				Properties of the observed FLAMES/GIRAFFE dataset in the Tucana dSph. Column (1) object-ID from fits header; (2) fiber-ID used through text; (3) RA-Dec coordinated in J2000; (4) V band magnitude transformed from DES photometric catalog (5) I band magnitude transformed from the same catalog (6) l.o.s. heliocentric velocity with error; (7) S/N ratio in pxl$^{-1}$ (the conversion factor to \AA$^{-1}$ is 2.24); (8) probability of membership according to the best-fitting kinematic model -- the two columns indicate the probabilities obtained using the FLAMES dataset alone ($P_1$), and combining with the FORS2 data($P_2$).}\\
			\hline\hline
			Object & Fiber & Ra, Dec (J2000) & $V$ & $I$ & $v_{hel} \pm \delta v_{hel}$ & SNR &  \multicolumn{2}{c}{P}  \\
			       &       & [deg]           &     &     & [km/s]                       & [pxl$^{-1}$] & P$_{1}$ & P$_{2}$  \\
			\hline
			\endfirsthead
			\caption{continued.}\\
			\hline\hline
			Object & Fiber & Ra, Dec (J2000) & $V$ & $I$ & $v_{hel} \pm \delta v_{hel}$ & SNR &  \multicolumn{2}{c}{P}  \\
			       &       & [deg]           &     &     & [km/s]                       & [pxl$^{-1}$] & P$_{1}$ & P$_{2}$  \\
			\hline
			\endhead
			\hline
			\endfoot
			 44068 &   1 & 340.34104, -64.366778  & 22.976 &  20.486 &    19.7 $\pm$   6.4 &  12.0 & Ph &    \\
    		 58039 &   2 & 340.27637, -64.447250  & 23.368 &  22.183 &               /     &   0.8 & Ph &    \\
    		 53427 &   3 & 340.38987, -64.422028  & 22.861 &  21.823 &               /     &   2.2 & Ph &    \\
    		 49196 &   4 & 340.40287, -64.401111  & 23.340 &  22.133 &               /     &   3.4 & Ph &    \\
    		100020 &   5 & 340.32892, -64.424889  & 24.880 &  22.406 &               /     &   3.3 & Ph &    \\
    		 51120 &   6 & 340.36892, -64.411639  & 22.996 &  21.890 &   160.6 $\pm$  18.8 &   3.2 & Ph &    \\
    		100019 &   7 & 340.38075, -64.453639  & 20.716 &  19.920 &    23.6 $\pm$   9.6 &  15.3 & Ph &    \\
    		 39027 &   8 & 340.54671, -64.333556  & 22.992 &  21.378 &    79.6 $\pm$  18.1 &   6.2 & 0.00 & 0.00  \\
    		 31104 &   9 & 340.34300, -64.279222  & 22.230 &  20.723 &   292.6 $\pm$  16.6 &   9.4 & 0.00 & 0.00  \\
    		 27260 &  10 & 340.47533, -64.253083  & 22.898 &  21.565 &    39.3 $\pm$   9.0 &   3.8 & Ph &    \\
    		 37852 &  11 & 340.49483, -64.326306  & 23.199 &  21.657 &     7.0 $\pm$   7.1 &   5.2 & 0.0 & 0.00   \\
    		 37244 &  12 & 340.42596, -64.322194  & 22.103 &  20.879 &   191.3 $\pm$  14.3 &   7.8 & 0.0 & 0.00   \\
    		 45927 &  13 & 340.13696, -64.379889  & 22.964 &  21.509 &   451.9 $\pm$   4.5 &   3.7 & Ph &    \\
    		 44431 &  14 & 340.31504, -64.368972  & 22.904 &  21.351 &  -201.6 $\pm$  13.2 &   4.9 & Ph &    \\
    		 45504 &  15 & 340.21796, -64.376917  & 23.023 &  21.510 &   234.5 $\pm$  13.8 &   5.4 & 0.00 & 0.00   \\
    		 40746 &  16 & 340.45733, -64.345333  & 22.190 &  21.000 &   230.3 $\pm$  18.6 &   6.6 & 0.00 & 0.00   \\
    		 30380 &  17 & 340.29579, -64.274167  & 23.089 &  21.909 &               /     &   3.5 & Ph &    \\
    		 34523 &  18 & 340.51400, -64.302361  & 22.602 &  21.179 &               /     &   5.3 & Ph &    \\
    		 32778 &  19 & 340.45058, -64.290778  & 24.225 &  21.797 &               /     &   3.8 & Ph &    \\
    		 32116 &  20 & 340.50783, -64.286083  & 22.204 &  20.595 &    32.4 $\pm$   5.2 &  11.8 & 0.00 & 0.00   \\
    		 43472 &  21 & 340.44033, -64.363083  & 22.615 &  21.174 &   -38.7 $\pm$  11.9 &   4.4 & Ph &    \\
    		 39444 &  22 & 340.25721, -64.336389  & 22.121 &  20.730 &   126.7 $\pm$   7.2 &  10.0 & 0.00 & 0.00   \\
    		 24956 &  23 & 340.37875, -64.235778  &  0.000 &   0.000 &               /     &   3.2 & Ph &    \\
    		 33846 &  24 & 340.36354, -64.297972  & 22.937 &  21.543 &   -42.5 $\pm$  10.3 &   5.3 & 0.00 & 0.00   \\
    		 42547 &  25 & 340.52042, -64.357194  & 22.577 &  21.207 &   105.9 $\pm$   8.6 &   4.9 & Ph &    \\
    		 30098 &  26 & 340.50979, -64.272278  & 17.807 &  16.986 &   123.9 $\pm$   1.1 & 109.6 & Ph &    \\
    		 20262 &  27 & 340.21804, -64.200806  & 22.595 &  20.826 &    36.0 $\pm$  18.1 &   8.6 & 0.00 & 0.00   \\
    		 28726 &  28 & 340.55079, -64.263167  & 23.994 &  21.447 &    16.6 $\pm$  18.2 &   5.7 & Ph &    \\
    		 28435 &  29 & 340.56392, -64.261222  &  0.000 &   0.000 &     9.2 $\pm$  13.2 &   3.8 & Ph &    \\
    		 36322 &  30 & 340.44375, -64.315444  & 22.305 &  20.703 &    30.3 $\pm$  11.8 &  10.2 & 0.00 & 0.00   \\
    		 40890 &  31 & 340.42958, -64.346389  & 23.053 &  21.507 &   -62.0 $\pm$  23.8 &   5.7 & 0.00 & 0.00   \\
    		 22014 &  32 & 340.56696, -64.213194  & 22.229 &  20.861 &   -22.4 $\pm$  10.0 &   7.9 & 0.00 & 0.00   \\
    		 24597 &  33 & 340.53050, -64.233250  & 23.047 &  21.985 &               /     &   2.4 & Ph &    \\
    		 20184 &  34 & 340.44471, -64.200333  & 22.705 &  20.813 &   119.3 $\pm$  12.0 &   6.8 & 0.00 & 0.00   \\
    		 25301 &  35 & 340.69692, -64.238194  & 23.488 &  22.373 &               /     &   3.1 & Ph &    \\
    		 21919 &  36 & 340.63321, -64.212528  & 22.609 &  21.043 &    61.9 $\pm$   7.7 &   6.1 & 0.00 & 0.00   \\
    		 31878 &  37 & 340.64842, -64.284611  & 22.696 &  21.652 &               /     &   4.3 & Ph &    \\
    		 24201 &  38 & 340.67017, -64.230417  & 22.557 &  21.386 &               /     &   2.9 & Ph &    \\
    		  4032 &  39 & 340.50846, -64.087278  & 22.940 &  21.389 &               /     &   1.5 & Ph &    \\
    		 16892 &  40 & 340.56733, -64.178917  & 22.773 &  21.559 &               /     &   1.9 & Ph &    \\
    		 31326 &  41 & 340.62975, -64.280972  & 22.065 &  20.503 &    12.7 $\pm$   7.2 &  13.4 & 0.00 & 0.00   \\
    		  5642 &  42 & 340.45183, -64.099333  &  0.000 &   0.000 &   169.0 $\pm$  14.0 &   2.2 & Ph &    \\
    		 36740 &  43 & 340.69904, -64.318389  & 22.802 &  21.257 &   226.6 $\pm$  13.3 &   5.1 & 0.00 & 0.00   \\
    		 13690 &  44 & 340.76192, -64.156333  & 22.987 &  21.887 &               /     &   2.1 & Ph &    \\
    		 32638 &  45 & 340.56775, -64.289861  & 22.388 &  20.678 &   111.2 $\pm$  14.0 &  12.0 & 0.00 & 0.00   \\
    		 32163 &  46 & 340.58575, -64.286389  & 22.559 &  20.710 &   117.7 $\pm$  10.7 &  10.4 & 0.00 & 0.00   \\
    		 30991 &  47 & 340.65567, -64.278278  & 18.715 &  17.675 &   -50.1 $\pm$   1.0 &  42.5 & Ph &    \\
    		 20980 &  48 & 340.70929, -64.205528  & 22.358 &  20.702 &     1.2 $\pm$   6.3 &  10.9 & 0.00 & 0.00   \\
    		 18416 &  49 & 340.65187, -64.188583  & 22.400 &  21.031 &    30.2 $\pm$   6.2 &   5.5 & 0.00 & 0.00   \\
    		 24848 &  50 & 340.61529, -64.235028  & 22.317 &  20.877 &   187.0 $\pm$  40.5 &   7.5 & Ph &    \\
    		 21528 &  51 & 340.70312, -64.209722  & 22.354 &  20.615 &    45.3 $\pm$   3.8 &   7.0 & 0.00 & 0.00   \\
    		 23752 &  52 & 340.47312, -64.226722  & 23.407 &  20.383 &   -23.1 $\pm$   8.4 &  10.2 & Ph &    \\
    		 31071 &  53 & 340.76042, -64.279000  & 22.194 &  20.633 &   -49.6 $\pm$   8.6 &  10.4 & 0.00 & 0.00   \\
    		 31001 &  54 & 340.52425, -64.278361  & 19.060 &  18.133 &    -8.5 $\pm$   1.3 &  48.2 & Ph &    \\
    		 16300 &  55 & 340.81579, -64.175000  & 22.232 &  20.746 &               /     &   6.2 & Ph &    \\
    		 19109 &  56 & 340.75412, -64.193778  & 23.158 &  21.607 &    10.0 $\pm$  28.4 &   3.2 & Ph &    \\
    		 18966 &  57 & 340.82242, -64.192750  & 22.928 &  21.727 &   -45.0 $\pm$  11.4 &   4.0 & Ph &    \\
    		 32477 &  58 & 340.72533, -64.288583  & 22.329 &  20.713 &   166.8 $\pm$   9.7 &   8.8 & 0.00 & 0.00   \\
    		 27434 &  59 & 340.42087, -64.254250  & 22.831 &  22.089 &               /     &   4.0 & Ph &    \\
    		 23221 &  60 & 340.79529, -64.222528  & 22.896 &  21.570 &               /     &   3.7 & Ph &    \\
    		 22156 &  61 & 340.92625, -64.214333  & 22.016 &  20.832 &    36.2 $\pm$   5.0 &   4.6 & Ph &    \\
    		 37211 &  62 & 340.60975, -64.322028  & 23.534 &  22.110 &    71.6 $\pm$   8.9 &   3.9 & Ph &    \\
    		 35540 &  63 & 340.60425, -64.309583  & 23.512 &  21.524 &    35.1 $\pm$   8.0 &   6.0 & 0.00 & 0.00   \\
    		 20829 &  64 & 340.86492, -64.204417  & 22.850 &  21.214 &    64.4 $\pm$   9.6 &   4.8 & Ph &    \\
    		 35289 &  65 & 340.95317, -64.307861  & 22.275 &  21.177 &   150.2 $\pm$  22.5 &   3.9 & Ph &    \\
    		 35409 &  66 & 340.89879, -64.308722  & 22.419 &  20.780 &   -53.4 $\pm$   5.6 &   7.1 & 0.00 & 0.00   \\
    		 32709 &  67 & 340.92150, -64.290306  & 22.070 &  20.786 &   145.1 $\pm$   5.7 &   8.4 & 0.00 & 0.00   \\
    		 28348 &  68 & 340.93237, -64.260694  & 21.678 &  21.149 &    74.5 $\pm$  15.9 &   1.2 & Ph &    \\
    		 21201 &  69 & 340.56067, -64.207028  & 22.672 &  20.824 &    21.8 $\pm$   7.2 &   6.3 & 0.00 & 0.00   \\
    		 34708 &  70 & 340.81462, -64.303722  & 22.380 &  21.111 &    31.0 $\pm$  10.6 &   5.1 & 0.00 & 0.00   \\
    		 26258 &  71 & 340.91354, -64.245583  & 22.960 &  21.634 &               /     &   4.3 & Ph &    \\
    		 40705 &  72 & 340.62008, -64.345167  & 23.701 &  22.204 &               /     &   4.1 & Ph &    \\
    		 25087 &  73 & 340.92037, -64.236667  & 23.166 &  21.664 &               /     &   3.2 & Ph &    \\
    		 43260 &  74 & 340.79483, -64.361667  & 21.816 &  20.598 &   119.6 $\pm$   6.4 &   7.1 & 0.00 & 0.00   \\
    		 38186 &  75 & 340.82146, -64.328167  & 22.477 &  21.367 &    -3.6 $\pm$  10.6 &   3.1 & Ph &    \\
    		 44653 &  76 & 340.78854, -64.370306  & 23.135 &  22.253 &   247.6 $\pm$  15.6 &   4.5 & Ph &    \\
    		 27730 &  77 & 340.57954, -64.256472  & 19.371 &  17.200 &    33.2 $\pm$   2.5 &  25.3 & Ph &    \\
    		 32288 &  78 & 340.60825, -64.287361  & 21.536 &  19.870 &     7.0 $\pm$   3.6 &  15.0 & Ph &    \\
    		 43618 &  79 & 340.65275, -64.363917  & 22.909 &  21.618 &    18.2 $\pm$  15.2 &   5.2 & 0.00 & 0.00   \\
    		 51984 &  80 & 340.82829, -64.415750  & 22.555 &  21.163 &    68.8 $\pm$   9.9 &   3.5 & Ph &    \\
    		 47042 &  81 & 340.80346, -64.387083  & 23.206 &  22.295 &    32.4 $\pm$  10.2 &   2.5 & Ph &    \\
    		 48384 &  82 & 340.67171, -64.396111  & 22.925 &  21.382 &   151.5 $\pm$  14.7 &   5.6 & 0.11 & 0.09   \\
    		 40996 &  83 & 340.55800, -64.347111  & 21.789 &  19.157 &     8.5 $\pm$   4.5 &  17.1 & Ph &    \\
    		 48246 &  84 & 340.90804, -64.395111  & 23.503 &  22.183 &    75.9 $\pm$  10.6 &   1.6 & Ph &    \\
    		  7717 &  85 & 340.36279, -64.114750  & 23.246 &  22.057 &  -127.7 $\pm$  16.4 &   2.0 & Ph &    \\
    		 53621 &  86 & 340.76875, -64.422778  & 22.723 &  21.204 &   138.3 $\pm$   7.6 &   4.5 & Ph &    \\
    		 46156 &  87 & 340.69479, -64.381472  & 23.079 &  21.690 &   -78.4 $\pm$  13.8 &   4.2 & Ph &    \\
    		100022 &  88 & 340.57396, -64.375972  & 21.449 &  20.089 &    -1.9 $\pm$   3.9 &  15.0 & 0.00 & 0.00   \\
    		 48450 &  89 & 340.69687, -64.396500  & 23.093 &  22.168 &    51.1 $\pm$  11.3 &   1.8 & Ph &    \\
    		 54894 &  90 & 340.66104, -64.428417  & 23.263 &  21.835 &   165.0 $\pm$  11.5 &   3.5 & Ph &    \\
    		100003 &  91 & 340.47454, -64.418306  &  0.000 &   0.000 &   178.5 $\pm$   3.6 &   8.5 & 0.99 & \tablefootmark{a}   \\
    		 49630 &  92 & 340.52779, -64.403528  & 23.248 &  22.008 &   173.9 $\pm$   9.5 &   3.2 & Ph &    \\
    		100014 &  93 & 340.45825, -64.394861  & 22.315 &  20.943 &   145.3 $\pm$   5.4 &   8.7 & 0.17 & \tablefootmark{a}   \\
    		100009 &  94 & 340.49612, -64.411722  &  0.000 &   0.000 &   188.7 $\pm$   3.6 &   8.6 & 0.99 & 0.99   \\
    		100013 &  95 & 340.50742, -64.407194  &  0.000 &   0.000 &    73.3 $\pm$  11.1 &   1.7 & Ph &    \\
    		100021 &  96 & 340.56921, -64.383500  & 22.354 &  20.657 &   186.2 $\pm$  12.6 &   6.8 & 0.80 & \tablefootmark{a}   \\
    		 52804 &  97 & 340.63075, -64.419389  & 22.454 &  21.103 &  -194.8 $\pm$   5.2 &   6.5 & 0.00 & 0.00   \\
    		 52613 &  98 & 340.61704, -64.418556  & 23.114 &  21.918 &   188.8 $\pm$  15.0 &   3.6 & Ph &    \\
    		 39325 &  99 & 340.55700, -64.335500  & 23.549 &  20.742 &     6.0 $\pm$  12.0 &   8.8 & Ph &    \\
    		100002 & 100 & 340.46446, -64.423806  & 21.317 &  20.260 &   188.0 $\pm$  15.6 &   5.7 & 0.99 & \tablefootmark{a}   \\
    		 49958 & 101 & 340.42304, -64.405278  & 22.881 &  21.748 &   167.9 $\pm$  12.7 &   3.4 & Ph &    \\
    		100016 & 102 & 340.40967, -64.402278  & 22.405 &  21.056 &   174.0 $\pm$   6.1 &   6.0 & 0.99 & \tablefootmark{a}   \\
    		100005 & 103 & 340.43446, -64.408333  & 22.420 &  21.021 &   180.9 $\pm$   5.0 &   8.0 & 0.99 & \tablefootmark{a}   \\
    		 50951 & 104 & 340.44454, -64.410722  & 22.730 &  21.691 &               /     &   2.8 & Ph &    \\
    		  9224 & 105 & 340.31279, -64.126778  & 22.600 &  21.255 &  -122.7 $\pm$  10.2 &   3.7 & Ph &    \\
    		100018 & 106 & 340.35333, -64.414500  & 22.687 &  20.941 &    67.1 $\pm$   3.9 &   7.1 & 0.00 & \tablefootmark{a}   \\
    		100017 & 107 & 340.39833, -64.411361  & 22.249 &  20.909 &   173.4 $\pm$   6.0 &   7.7 & 0.99 & \tablefootmark{a}   \\
    		100006 & 108 & 340.49225, -64.423611  & 22.349 &  21.126 &   191.5 $\pm$   8.2 &   5.6 & 0.99 & \tablefootmark{a}   \\
    		100001 & 109 & 340.44696, -64.422306  & 21.427 &  20.323 &   175.1 $\pm$   3.8 &   8.4 & 0.99 & \tablefootmark{a}   \\
    		 47569 & 110 & 340.34308, -64.390611  & 22.639 &  21.297 &   149.9 $\pm$  11.1 &   5.3 & 0.63 & 0.54   \\
    		 46004 & 111 & 340.42396, -64.380417  & 22.728 &  21.480 &    18.6 $\pm$  17.8 &   5.0 & 0.00 & 0.00   \\
    		 54972 & 112 & 340.25037, -64.428806  & 23.127 &  22.017 &               /     &   2.2 & Ph &    \\
    		 43720 & 113 & 340.41837, -64.364556  & 21.006 &  19.251 &    63.2 $\pm$   2.5 &  30.8 & Ph &    \\
    		 21138 & 114 & 340.59517, -64.206583  & 23.023 &  21.318 &   105.6 $\pm$  26.0 &   6.0 & 0.00 & 0.00   \\
    		 39993 & 115 & 340.32687, -64.340028  & 21.844 &  20.614 &   -43.2 $\pm$  10.5 &   6.5 & 0.00 & 0.00   \\
    		 50864 & 116 & 340.38775, -64.410250  & 22.784 &  21.569 &               /     &   3.0 & Ph &    \\
    		 39557 & 117 & 340.14258, -64.337139  & 22.481 &  21.281 &               /     &   2.7 & Ph &    \\
    		 25558 & 118 & 340.42058, -64.240028  & 22.349 &  20.730 &   172.8 $\pm$   8.6 &   5.9 & 0.00 & 0.00   \\
    		 33805 & 119 & 340.28633, -64.297639  & 22.541 &  21.346 &  -364.1 $\pm$   7.5 &   3.3 & Ph &    \\
    		 34818 & 120 & 340.29067, -64.304556  & 22.643 &  21.338 &               /     &   3.0 & Ph &    \\
    		 28959 & 121 & 340.13192, -64.264889  & 22.808 &  21.283 &               /     &   2.5 & Ph &    \\
    		 26205 & 122 & 340.18454, -64.245194  & 22.136 &  20.540 &    -1.2 $\pm$   6.2 &   5.3 & 0.00 & 0.00   \\
    		 32371 & 123 & 340.25575, -64.287861  & 22.674 &  20.963 &   161.0 $\pm$  13.4 &   4.5 & Ph &    \\
    		 26562 & 124 & 340.47250, -64.248028  & 22.589 &  21.159 &  -404.5 $\pm$   8.8 &   4.3 & Ph &    \\
    		 33199 & 125 & 340.05833, -64.293333  & 23.122 &  22.000 &               /     &   2.7 & Ph &    \\
    		 38665 & 126 & 340.39025, -64.331333  & 20.370 &  19.311 &   100.5 $\pm$   5.4 &  16.8 & Ph &    \\
    		 15314 & 127 & 340.19367, -64.168111  & 22.785 &  21.271 &   193.8 $\pm$  23.0 &   2.6 & Ph &    \\
    		 25202 & 128 & 340.42783, -64.237500  & 23.155 &  21.857 &   319.1 $\pm$  12.4 &   2.7 & Ph &    \\
    		 22526 & 129 & 340.27612, -64.217083  & 22.241 &  19.161 &     6.8 $\pm$   7.8 &  11.0 & Ph &    \\
    		 28804 & 130 & 340.29137, -64.263722  & 22.447 &  21.320 &  -419.7 $\pm$   3.3 &   3.0 & Ph &    \\
    		 22297 & 131 & 340.13142, -64.215389  & 21.895 &  20.688 &   -37.1 $\pm$  10.1 &   5.1 & 0.00 & 0.00   \\
    		 25382 & 132 & 340.18987, -64.238778  &  0.000 &   0.000 &               /     &   3.0 & Ph &    \\
    		 13504 & 133 & 340.24646, -64.155139  & 22.938 &  21.329 &   136.9 $\pm$  15.8 &   3.6 & Ph &    \\
    		 19228 & 134 & 340.57158, -64.194472  & 23.355 &  22.151 &               /     &   2.9 & Ph &    \\
    		  5411 & 135 & 340.47821, -64.097583  & 22.284 &  21.120 &               /     &   3.0 & Ph &    \\
    		  7674 & 136 & 340.45483, -64.114306  & 22.383 &  21.151 &               /     &   3.2 & Ph &    \\
    		 24957 & 137 & 340.75162, -64.235778  & 22.854 &  21.301 &               /     &   4.2 & Ph &    \\
    		 24991 & 138 & 340.76054, -64.235972  & 23.274 &  22.108 &               /     &   2.6 & Ph &    \\
    		 31905 & 139 & 340.54608, -64.284778  & 22.730 &  21.167 &    45.5 $\pm$  16.0 &   5.6 & 0.00 & 0.00   \\
    		 20561 & 140 & 340.84067, -64.202861  & 21.969 &  20.723 &   199.4 $\pm$  10.7 &   6.1 & 0.00 & 0.00   \\
    		 13652 & 141 & 340.89246, -64.156111  & 22.599 &  20.829 &               /     &   3.2 & Ph &    \\
    		 29029 & 142 & 340.98737, -64.265417  & 23.359 &  22.117 &               /     &   1.9 & Ph &    \\
    		 28008 & 143 & 340.92996, -64.258472  &  0.000 &   0.000 &    83.4 $\pm$  14.9 &   4.8 & Ph &    \\
    		 43939 & 144 & 340.83217, -64.366000  & 22.510 &  20.554 &    40.1 $\pm$   5.4 &   7.1 & 0.00 & 0.00   \\
    		 41813 & 145 & 340.53267, -64.352583  & 22.750 &  21.276 &   142.6 $\pm$  18.3 &   6.3 & 0.01 & 0.01   \\
    		 40477 & 146 & 340.93167, -64.343667  & 19.724 &  18.922 &   168.6 $\pm$   3.4 &  19.6 & Ph &    \\
    		 38363 & 147 & 340.56947, -64.329389  & 23.588 &  21.699 &   -44.2 $\pm$   9.7 &   3.2 & Ph &    \\
    		 54686 & 148 & 340.83783, -64.427528  & 22.108 &  20.905 &   180.1 $\pm$  11.6 &   3.4 & Ph &    \\
    		 41818 & 149 & 340.48062, -64.352611  &  0.000 &   0.000 &   160.9 $\pm$  21.5 &   3.5 & Ph &    \\
    		 52574 & 150 & 340.77154, -64.418333  & 23.343 &  22.174 &    68.4 $\pm$   5.3 &   2.0 & Ph &    \\
    		 50194 & 151 & 340.45900, -64.406667  & 23.010 &  21.531 &               /     &   3.8 & Ph &    \\
    		 52007 & 152 & 340.54187, -64.415889  & 23.029 &  21.919 &               /     &   2.8 & Ph &    \\
    		100015 & 153 & 340.51367, -64.411667  & 22.436 &  20.912 &   181.9 $\pm$   6.3 &   5.2 & 0.99 & \tablefootmark{a}   \\
    		 52859 & 154 & 340.60392, -64.419611  & 22.947 &  21.748 &               /     &   2.3 & Ph &    \\
    		 54691 & 155 & 340.63350, -64.427556  & 23.203 &  22.063 &               /     &   2.5 & Ph &    \\
    		 60765 & 156 & 340.69737, -64.465083  & 22.639 &  21.274 &   147.8 $\pm$  17.3 &   3.0 & Ph &    \\
    		 50636 & 157 & 340.61592, -64.409111  & 23.031 &  21.272 &               /     &   3.2 & Ph &    \\
    		 50350 & 158 & 340.45050, -64.407500  & 22.775 &  21.648 &   197.8 $\pm$   9.3 &   2.6 & Ph &    \\
    		 46819 & 159 & 340.41362, -64.385528  & 22.469 &  20.773 &   -34.9 $\pm$  11.4 &   6.6 & 0.00 & 0.00   \\
    		100007 & 160 & 340.42029, -64.416056  & 22.412 &  21.129 &   -88.3 $\pm$   6.2 &   4.7 & Ph &    \\
    		 53700 & 161 & 340.46612, -64.423111  &  0.000 &   0.000 &   185.6 $\pm$  11.5 &   2.5 & Ph &    \\
    		 51422 & 162 & 340.36446, -64.413167  & 23.310 &  22.180 &   204.1 $\pm$   4.6 &   2.8 & Ph &    \\
    		 51322 & 163 & 340.18804, -64.412694  & 22.476 &  21.120 &             /       &   3.0 & Ph &    \\
     		 43254 & 164 & 340.23679, -64.361611  & 22.178 &  21.031 &   229.1 $\pm$   9.0 &   2.0 & Ph &    \\
			\hline    
		\end{longtable}
		\tablefoot{In the P columns, stars marked as "Ph" are non-members excluded according to their magnitudes and colors or because their measurements resulted to be not reliable or because having a S/N$<10$\AA$^{-1}$ (see Sect.~\ref{sec:membership} and Appendix~\ref{sec:A} for full description). 
		\tablefoottext{a}{In common with the FORS2 dataset}.}
	}
	
	\begin{appendix}
	
	\section{L.o.s. velocity and metallicity measurements -- Consistency checks and comparisons}\label{sec:A}
	
	\subsection{Consistency checks}\label{subsec:sanity-vel}
	We have performed consistency checks in order to unveil the presence of possible systematic errors in our velocity determinations using \textit{fxcor}, or other issues.
	
	For all the datasets analyzed, we searched for systematics by shifting the template spectrum at several velocities (from -50 km\,s$^{-1}$ to 500 km\,s$^{-1}$ at step of 50 km\,s$^{-1}$) and cross-correlating them with the templates themselves at rest, through the full wavelength range of data and also only around the CaT. We did not find any significant systematic shift introduced by the cross-correlation procedure as a function of the velocity shift nor the cross-correlation range.
	
	We also verified if the random errors are well treated by the \textit{fxcor} task and if they lead to reliable velocity errors. In this case we performed a Monte Carlo analysis, randomly adding to the shifted templates of the previous step the error-spectra from targets having measured S/N of 10, 15, 20, 25 and 30 \AA$^{-1}$ for the FORS2 datasets, and of 5, 10, 20, 50 and 100 \AA$^{-1}$ for the FLAMES one. We performed 250 realizations for each case and obtained that the procedure tends to produce velocity errors that are underestimated with respect to the velocity scatter from the individual Monte Carlo runs for S/N lower than 10 \AA$^{-1}$, for both datasets. Therefore we take these values as the limit S/N above which we will trust the velocity error estimations. At this S/N the velocity errors resulted to be $\sim \pm10$ km s$^{-1}$\, in both cases. 
	
	We further checked the choice of using the two reddest lines of the CaT for the cross-correlation measurements of the FLAMES dataset (see previous section), creating two more catalogs of velocity estimations using both the entire CaT range and the two bluest lines. Cross-correlating between catalogs and selecting those targets having a S/N $>10$ \AA$^{-1}$, which reduced our sample to 76 objects out of 164, we found very good agreement, except for only 8 targets that had velocity measurements differences at more than 3-$\sigma$ from each others. We visually checked the corresponding spectra for these targets and found that for the outliers the first and/or third line of the CaT resulted affected by sky residuals or were hidden by noise, while in general the central line resulted more clearly visible. For fiber 118 happened the opposite and we considered for this case the measurement obtained using the full CaT range.
	
	Considering again the velocities from targets having a S/N $>10$ \AA$^{-1}$, we looked for cases were the cross-correlation procedure did not get a velocity solution, finding only two cases (fibers 18 and 55). Their visual inspection did not show obvious CaT lines, and we thus decided to discard these two targets.
	
	Finally we checked the internal accuracy of our P91 FORS2 dataset by inspecting the four targets in common between the two pointings.  Of these, three had reliable measurements of velocity and metallicity which were compatible within 1-$\sigma$, while the other did not have reliable measurements. In fact it did not show a visible CaT and its \textit{I}-magnitude and color were also not compatible with the RGB of Tucana (see also Sect.\ref{sec:membership}). We decided then to discard this target from the sample and to average together the measurements from the other three targets.
	
	We further excluded two more targets from the P91 FORS2 dataset since their measurements were not reliable due to high residuals in their spectra: these are slit 14, chip-2 of Tuc0 field and slit 17, chip-1 of Tuc1. The high residuals in these two spectra were due to the stellar trace falling on a bad CCD-row and a poor sky-subtraction that could not be improved, respectively.

	\subsection{Comparison between measurements of different datasets obtained with our reduction}\label{subsec:comparison-flames}

	We compared the measurements obtained from the P91 dataset with those from our reduction of the P69 and FLAMES datasets, using the targets they had in common. This was an important step to ensure we could combine the velocity catalogues.
	
\begin{figure*}
		\centering
		\includegraphics[width=\hsize]{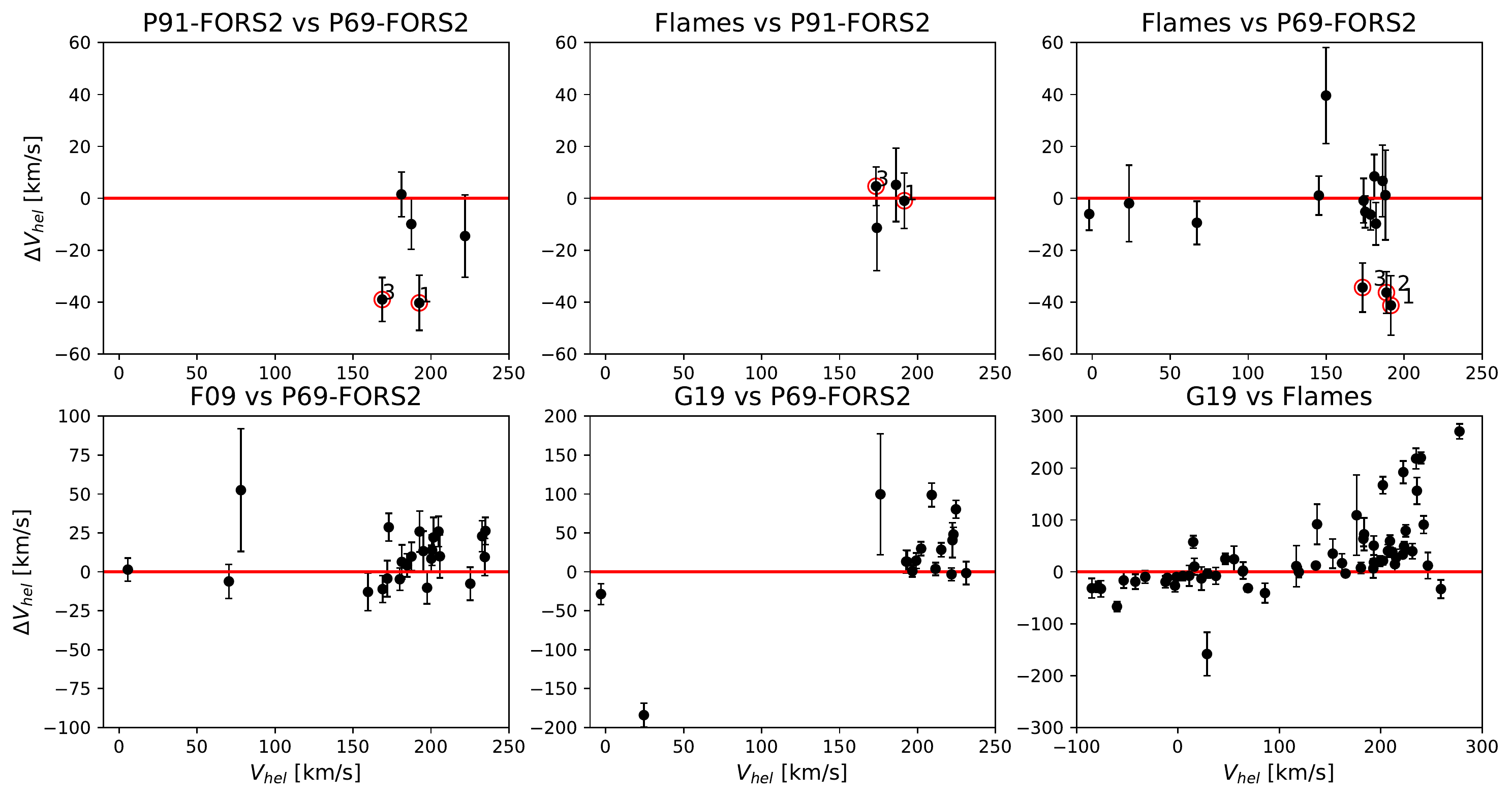}
		\caption{Velocity comparison for the common targets between datasets -- v$_{hel}$ vs. $\Delta$v$_{hel}$.
			\textit{Top:} between the datasets from our internal reduction -- \textit{left}: the FORS2 datasets; \textit{central}: Flames vs. P91-FORS2; \textit{right}: Flames vs. P69-FORS2. Red circles indicate the problematic targets highlighted in the main text.
			\textit{Bottom:} between our datasets and those from the literature -- \textit{left}: F09 vs. P69-FORS2; \textit{central}: G19 vs. P69-FORS2; \textit{right}: G19 vs. Flames.
		}
		\label{Fig:A.1}
	\end{figure*}

	The P91 and P69 FORS2 datasets have 5 targets in common with measured velocities and metallicities, while the FLAMES dataset has in common 7 targets (4 taking those with S/N $>10$ \AA$^{-1}$) with the P91 FORS2 dataset and 18 (15) with the P69 one (see the top row of Fig.~\ref{Fig:A.1} for the velocity comparison). 
	
	The agreement between these common targets is excellent, mostly within 1-$\sigma$, except for three targets in common between the FLAMES dataset and the P69 FORS2 one (marked as object 1, 2 and 3 in the top row of Fig.~\ref{Fig:A.1}) with velocities at more than 3-$\sigma$  difference from each others. Of these targets, two are in common with the P91 FORS2 dataset (objects 1 and 3), which are in good agreement with the FLAMES measurements (Fig.~\ref{Fig:A.1} top row, central panel), but not with those of the P69 FORS2 dataset  (Fig.~\ref{Fig:A.1} top row, right panel). We found then that the problem may reside with the measurements from the P69 FORS2 catalog. Therefore we inspected these outliers, namely slits 9 and 10 of chip-1 and slit 13 of chip-2 (respectively object 1, 2 and 3 in the cited figures), finding that slits 9 and 13 have particularly noisy sky-lines residual around the first two CaT lines that may have compromised their measurements, while the spectrum of slit 10 was clean. However this target was marked as a double star and with slit mis-centering problems in F09. We decided then to exclude these values from the P69 FORS2 catalog, but maintaining them in the P91 FORS2 and FLAMES datasets.
	
	In conclusion, after a homogeneous analysis, the measurements for stars in common, which were taken during several epochs as well as with different instruments and spectral resolution in the case of FORS2 and FLAMES, are found to be in very good agreement.
	
	\subsection{Comparison with other works} \label{subsec:vel-comparison}
	
	{\it Comparison with F09:} The l.o.s. velocity and metallicity distributions of the P91 and F09 datasets show a difference of $\sim10$ km\,s$^{-1}$ and $\sim0.25$ dex in the mean values, respectively. We investigated whether clear offsets could be found by comparing the measurements of the targets in common between the two studies. There are only three such stars, for which the metallicities agree at the 1-$\sigma$ level, but the l.o.s. velocities differ at 3-$\sigma$ level for two stars and 1.5-$\sigma$ for the other one, respectively. The stars discrepant at the 3-$\sigma$ level were slits 9 and 13 of chip-1 and -2, respectively (targets 7 and 18 in F09), whose P91 spectra do not show any particular issue. 
	
	Since we could not pin point the sources of these differences in the line-of-sight velocities, we decided to re-reduce and analyze again the F09 FORS2 dataset, so to be homogeneous with the treatment of the P91 data. 
	
	The comparison of the l.o.s. velocities and metallicities derived from our reduction of the P69 FORS2 dataset and those published by F09 showed average offsets of 7 km\,s$^{-1}$ and 0.2 dex in the velocity and metallicity measurements, of the same signs and comparable to those previously found between the targets in common between the P91 FORS2 dataset and the F09 catalog (see Fig.~\ref{Fig:A.1} bottom row, left panel for the velocity comparison). 
	It is unclear what leads to the measured difference in velocity. For the metallicities, the [Fe/H]-CaT EW calibration in F09 is equivalent to that used here in the metallicity and magnitude regime of Tucana's stars, and the 0.1~mag difference in the adopted $V_{\rm HB}$ value does not have a significant impact on the final [Fe/H] values. The main source of difference is likely to reside in how the EWs were calculated: in F09 Gaussian profiles were used to fit the CaT lines, while we adopt Voigt profiles. We checked that if we were to adopt Gaussian profiles, we would obtain lower metallicities of a difference comparable to the systematic shift we previously found. 
	We want to stress, however, that the Voigt profiles represent a better fit to the CaT lines, in particular along the line-wings (see e.g. \citealp{Starkenburg2010}).		

	{\it Comparison with G19:} We compared our own reduced datasets with the full catalog of observed targets by G19. 
	There was not a good agreement in all cases. The P91 FORS2 dataset was the one with the lowest number of common targets (6), which showed mainly a systematic offset of $\sim30$ km\,s$^{-1}$. One target, however, showed a velocity difference of $\sim130$ km\,s$^{-1}$. We visually checked for it in our dataset (target 13 of chip-1 in Tuc0), finding a clean high S/N spectrum with no features that could have biased the velocity measurement. 
	
	The P69 FORS2 dataset resulted  to have 16 targets in common with G19 (see Fig.~\ref{Fig:A.1} bottom row, central panel). While for a few targets there is good agreement, overall there is a large scatter in the distribution of velocity differences with one significant outlier (target 1 of chip-2). An inspection of its corresponding spectrum in the P69 data showed the CaT lines shifted with respect to the rest frame by $\sim 200$ km\,s$^{-1}$, compatible with what we found in the cross-correlation measurements, but significantly different from the $\sim25 \pm 14$ km\,s$^{-1}$ reported by G19. 
	
	The comparison between the velocities we derived from our treatment of the FLAMES dataset was still more puzzling, considering that the values obtained come from the same sample. As can be seen from the bottom right panel of Fig.~\ref{Fig:A.1}, there is a significant scatter among the 59 targets in common (we compared only those measurements from our dataset with a S/N $>10$ \AA$^{-1}$), in particular at high velocities ($>100$ km\,s$^{-1}$), and there might even be a dependence as a function of velocity between the two datasets. We also confirmed that these problems remain when considering only those targets marked as likely members in G19. 
	
	G19 find a systemic velocity for Tucana around 200 km\,s$^{-1}$. At $v_{\rm hel, G19} \sim 200$ km\,s$^{-1}$, the comparison in Fig.~\ref{Fig:A.1} (bottom row, right panel) gives velocities around 170-190 km\,s$^{-1}$ for our FLAMES measurements of the same stars (i.e. velocities close to the systemic velocity we find for Tucana), as well as values between 0-50 km\,s$^{-1}$, i.e. typical of foreground stars. 
	Our suspicion is that the G19 catalog contains an excess of stars with velocities around 200 km\,s$^{-1}$.
	
	We investigate this possibility in two ways: by comparing the radial distribution of the G19 members with that obtained from photometric observations of Tucana's stars (test A) and comparing the G19 l.o.s. velocity distribution for the stars they considered as non-members with the expectations from a Galactic foreground model (test B). In both cases, we took into account the displacement of the FLAMES pointings from the center of Tucana. 
	
	{\it Test A} We looked at the normalized cumulative radial distribution of the G19's member stars, and compared it to that obtained from the observed surface density profile of Tucana that we measured from the VLT/VIMOS photometric catalog introduced in Sect.\ref{sec:membership}. 
	
	Although the VLT/VIMOS catalog does not cover the entire area of the FLAMES pointings, it was sufficient for following the surface density profile of the RGB stars of Tucana up to its nominal tidal radius. 
	
	We found that the number of G19's member stars in the outer parts of Tucana, up to where VLT/VIMOS photometry extends, tends to be overestimated with respect to that expected from the photometry for Tucana's RGB stars in the same area. 
	
	It could be however argued that the FLAMES/GIRAFFE fiber set-up might imprint a different distribution than that expected from the photometry. Therefore, we perform the following {\it Test B}, which is instead free from this possible issue. 
	
	{\it Test B} We now concentrate on the l.o.s. velocity distribution of the stars marked as non-members in G19, which should only contain Galactic contaminants: this distribution shows two clear peaks in the velocity histogram, one around 0 km\,s$^{-1}$ and, unexpectedly, the other at $\sim200$ km\,s$^{-1}$ (G19, Fig.~6). Making the reasonable assumption that the non-members are mainly foreground Galactic contaminants, we compared their velocity distribution with that obtained from the Besan\c{c}on model \citep{Robin2003} generated in the direction of Tucana over an area equivalent to that of a FLAMES-GIRAFFE pointing and by selecting the Besan\c{c}on model stars to have similar position on the CMD as the FLAMES targets. 
	The velocity distribution of the Besan\c{c}on model stars showed just a single peak around 0 km\,s$^{-1}$, with a smooth decline towards negative and positive velocities, with a tail extending to 300 km\,s$^{-1}$. We checked if the peak at $\sim200$ km\,s$^{-1}$ for the G19's non-members could be explained from the distribution expected from the Galactic model. We randomly chose from the synthetic dataset a number of stars equal to that of the G19 non-members and, over 1000 trials, calculated the number of objects extracted from the Besan\c{c}on model which would have velocities $>150$ km\,s$^{-1}$: we never got a number of contaminants as high as that of G19's non-members over the same velocity range. This indicates that the number of stars with velocities $\sim200$ km\,s$^{-1}$ in the G19 catalog of non-members could likely be overestimated. 
	Performing the same exercise for our FLAMES targets that resulted to be non-members (i.e. having $P<0.05$) considering those targets with velocities $>120$ km\,s$^{-1}$ (accounting for the observed shift between the datasets), we found that the observed number of non-members is within the 87\% (1.5-$\sigma$) of the distribution obtained for the contaminants.
	
	We speculate that this excess of velocities around $\sim$200 km\,s$^{-1}$ in the G19 dataset can be attributed to a sky subtraction problem around the CaT region. In fact, the sky-lines at 8504\AA\ and 8548\AA, which are around the first and second lines of the CaT, if badly  subtracted could lead to strong absorption residuals. These features, in a low-S/N regime and during the cross-correlation procedure, could be mistaken for the first two CaT lines shifted at $\sim 200$ km\,s$^{-1}$, i.e. around the value of the systemic velocity reported for Tucana by G19. 
	
	\section{Supplementary material from the kinematic analysis}\label{sec:B}
	\subsection{Supplementary tables}\label{sec:B.1}
	In all tables, the reported values of the parameters obtained from a Bayesian analysis represent the median of the corresponding marginalized posterior distributions, with 1-$\sigma$ errors set as the confidence intervals around the central value enclosing $68 \%$ of each distributions.
	
	\begin{table*}
		\caption{Structural parameters from the MCMC analysis fitting an exponential density profile to the RGB population selected in the VIMOS photometry. We report the value of the central density, the coordinates of the optical center, the scale parameters projected along the optical major axis, the position angle of the optical major axis, the ellipticity and the density of constant contamination.}             
		\label{table:struct-params}      
		\centering          
		\begin{tabular}{c c c c c c}
			\hline\hline
			$\sigma_0$ & ($\alpha_0,\delta_0$) & $r_0$ & P.A. & $\epsilon$  & $\sigma_c$ \\ 
			${\rm (stars\,arcmin^{-2})}$ & (deg) & (arcmin) & (deg) &  & ${\rm (stars\,arcmin^{-2})}$\\ 
			\hline
			$475_{-30}^{+32}$ & $340.4589 \pm 0.0006$, $-64.4198 \pm 0.0004$ & $0.72 \pm 0.03$ & $95.6 \pm 1.9$ & $0.46 \pm 0.02$ & $1.1 \pm 0.2$ \\
			\hline
		\end{tabular}
	\end{table*}
	
	\begin{table*}
		\centering
		\caption{Mock results for the P91+P69+FLAMES ($P>0.05$) catalog using an input linear rotation model. The first two columns represent the input parameters of the rotation model. In all cases the input systemic velocity V$_{\rm sys}$ and velocity dispersion $\sigma_v$ were fixed to 0 and 6 km\,s$^{-1}$, respectively. The middle columns are the recovered parameter fitting a linear rotation model and a dispersion-only one, respectively. The last column is the Bayes factor accounting for the evidences of the two models.} 
		\label{table:mock}
		\begin{tabular}{cc|rrrr|rr|r}
			\hline\hline
			\textit{k} & $\theta_k$ & V$_{\rm sys}$ & $\sigma_v$ & \textit{k} & $\theta_k$ & V$_{\rm sys}$ & $\sigma_v$ & $\textrm{ln}B_{\rm rot,disp}$ \\ 
			${\rm (km/s/')}$& ($^\circ$) & (km/s) & (km/s) & (km/s/$'$) & ($^\circ$) & (km/s) & (km/s) & \\ \hline 
			& 97 & $0.03_{-1.22}^{+1.26}$ & $6.08_{-1.26}^{+1.11}$ & $11.40_{-0.77}^{+0.88}$ & $96.83_{-9.18}^{+9.17}$ & $2.69_{-1.25}^{+1.22}$ & $16.56_{-1.17}^{+1.33}$ & $34.06_{-4.78}^{+5.43}$ \\
			11.2 & 142 & $0.00_{-1.27}^{+1.18}$ & $6.03_{-1.18}^{+1.17}$ & $11.10_{-1.58}^{+1.61}$ & $141.36_{-6.81}^{+6.05}$ & $0.20_{-1.27}^{+1.11}$ & $12.05_{-1.24}^{+1.30}$ & $19.67_{-4.55}^{+5.36}$ \\
			& 187 & $-0.07_{-1.29}^{+1.24}$ & $6.12_{-1.36}^{+1.14}$ & $10.84_{-1.80}^{+2.01}$ & $187.55_{-4.70}^{+5.55}$ & $-2.36_{-1.23}^{+1.16}$ & $9.70_{-1.21}^{+1.14}$ & $9.92_{-3.97}^{+4.49}$ \\ \hline
			& 97 & $0.08_{-1.28}^{+1.17}$ & $6.00_{-1.30}^{+1.11}$ & $7.73_{-0.87}^{+0.84}$ & $96.03_{-13.37}^{+12.62}$ & $1.76_{-1.28}^{+1.09}$ & $11.63_{-1.26}^{+1.19}$ & $19.11_{-4.55}^{+4.84}$ \\
			7.5 & 142 & $0.00_{-1.13}^{+1.08}$ & $5.97_{-1.26}^{+1.17}$ & $7.35_{-1.57}^{+1.75}$ & $141.11_{-12.74}^{+9.21}$ & $0.06_{-1.03}^{+1.13}$ & $9.06_{-1.28}^{+1.18}$ & $9.26_{-4.02}^{+4.43}$ \\
			& 187 & $-0.11_{-1.18}^{+1.12}$ & $5.98_{-1.26}^{+1.21}$ & $7.01_{-1.91}^{+1.92}$ & $187.18_{-7.34}^{+8.22}$ & $-1.55_{-1.11}^{+1.06}$ & $7.85_{-1.24}^{+1.16}$ & $3.12_{-2.90}^{+3.56}$ \\\hline
			& 97 & $0.01_{-1.21}^{+1.16}$ & $6.10_{-1.23}^{+1.14}$ & $5.94_{-0.95}^{+0.91}$ & $96.35_{-17.49}^{+16.68}$ & $1.19_{-1.10}^{+1.15}$ & $9.59_{-1.31}^{+1.12}$ & $10.85_{-3.87}^{+4.75}$ \\
			5.6 & 142 & $-0.03_{-1.22}^{+1.30}$ & $6.01_{-1.21}^{+1.14}$ & $5.46_{-1.54}^{+1.47}$ & $139.47_{-16.20}^{+11.33}$ & $0.04_{-1.14}^{+1.26}$ & $7.88_{-1.20}^{+1.10}$ & $4.41_{-3.29}^{+3.54}$ \\
			& 187 & $-0.17_{-1.25}^{+1.22}$ & $6.10_{-1.25}^{+1.16}$ & $5.03_{-2.15}^{+1.97}$ & $188.44_{-10.14}^{+12.05}$ & $-1.19_{-1.18}^{+1.14}$ & $7.18_{-1.25}^{+1.13}$ & $0.41_{-2.30}^{+2.92}$ \\\hline
			& 97 & $0.04_{-1.21}^{+1.17}$ & $6.07_{-1.28}^{+1.12}$ & $4.08_{-0.92}^{+0.95}$ & $94.80_{-23.97}^{+25.01}$ & $0.81_{-1.13}^{+1.07}$ & $7.67_{-1.15}^{+1.20}$ & $3.85_{-3.13}^{+3.42}$ \\
			3.7 & 142 & $-0.10_{-1.18}^{+1.26}$ & $6.03_{-1.23}^{+1.16}$ & $3.50_{-1.27}^{+1.46}$ & $139.67_{-27.83}^{+17.43}$ & $0.01_{-1.15}^{+1.17}$ & $6.88_{-1.20}^{+1.27}$ & $0.15_{-1.99}^{+3.04}$ \\
			& 187 & $-0.13_{-1.23}^{+1.13}$ & $6.00_{-1.19}^{+1.17}$ & $2.87_{-1.58}^{+2.10}$ & $188.08_{-15.21}^{+18.70}$ & $-0.73_{-1.17}^{+1.06}$ & $6.55_{-1.21}^{+1.17}$ & $-1.68_{-1.24}^{+1.98}$ \\\hline
			& 97 & $0.07_{-1.23}^{+1.12}$ & $6.10_{-1.31}^{+1.06}$ & $2.25_{-1.06}^{+1.11}$ & $97.21_{-50.07}^{+45.31}$ & $0.47_{-1.21}^{+1.06}$ & $6.56_{-1.24}^{+1.18}$ & $-1.40_{-1.46}^{+2.48}$ \\
			1.9 & 142 & $-0.08_{-1.10}^{+1.13}$ & $6.00_{-1.28}^{+1.14}$ & $1.63_{-1.23}^{+1.47}$ & $152.44_{-46.04}^{+26.47}$ & $-0.04_{-1.13}^{+1.08}$ & $6.21_{-1.14}^{+1.17}$ & $-2.41_{-0.79}^{+1.53}$ \\
			& 187 & $-0.05_{-1.14}^{+1.09}$ & $6.01_{-1.35}^{+1.10}$ & $1.26_{-1.13}^{+1.74}$ & $183.09_{-27.01}^{+31.00}$ & $-0.32_{-1.11}^{+1.05}$ & $6.16_{-1.17}^{+1.11}$ & $-2.81_{-0.52}^{+1.42}$ \\\hline
			& 97 & $0.01_{-1.16}^{+1.10}$ & $5.97_{-1.17}^{+1.19}$ & $0.17_{-1.48}^{+1.18}$ & $146.72_{-116.54}^{+30.94}$ & $0.01_{-1.12}^{+1.07}$ & $6.03_{-1.16}^{+1.15}$ & $-3.13_{-0.28}^{+0.80}$ \\
			0.0 & 142 & $-0.02_{-1.18}^{+1.09}$ & $5.97_{-1.24}^{+1.06}$ & $-0.18_{-1.24}^{+1.52}$ & $168.49_{-48.65}^{+29.45}$ & $-0.03_{-1.16}^{+1.05}$ & $6.02_{-1.23}^{+1.12}$ & $-3.09_{-0.33}^{+0.89}$ \\
			& 187 & $0.01_{-1.20}^{+1.20}$ & $5.91_{-1.24}^{+1.16}$ & $-0.04_{-1.34}^{+1.38}$ & $176.81_{-30.81}^{+42.76}$ & $0.02_{-1.20}^{+1.14}$ & $5.98_{-1.27}^{+1.16}$ & $-3.11_{-0.31}^{+0.85}$ \\\hline
		\end{tabular}
	\end{table*}
	
	\subsection{CDM halo profiles}\label{sec:C}
	
\begin{figure}
		\centering
		\includegraphics[width=\hsize]{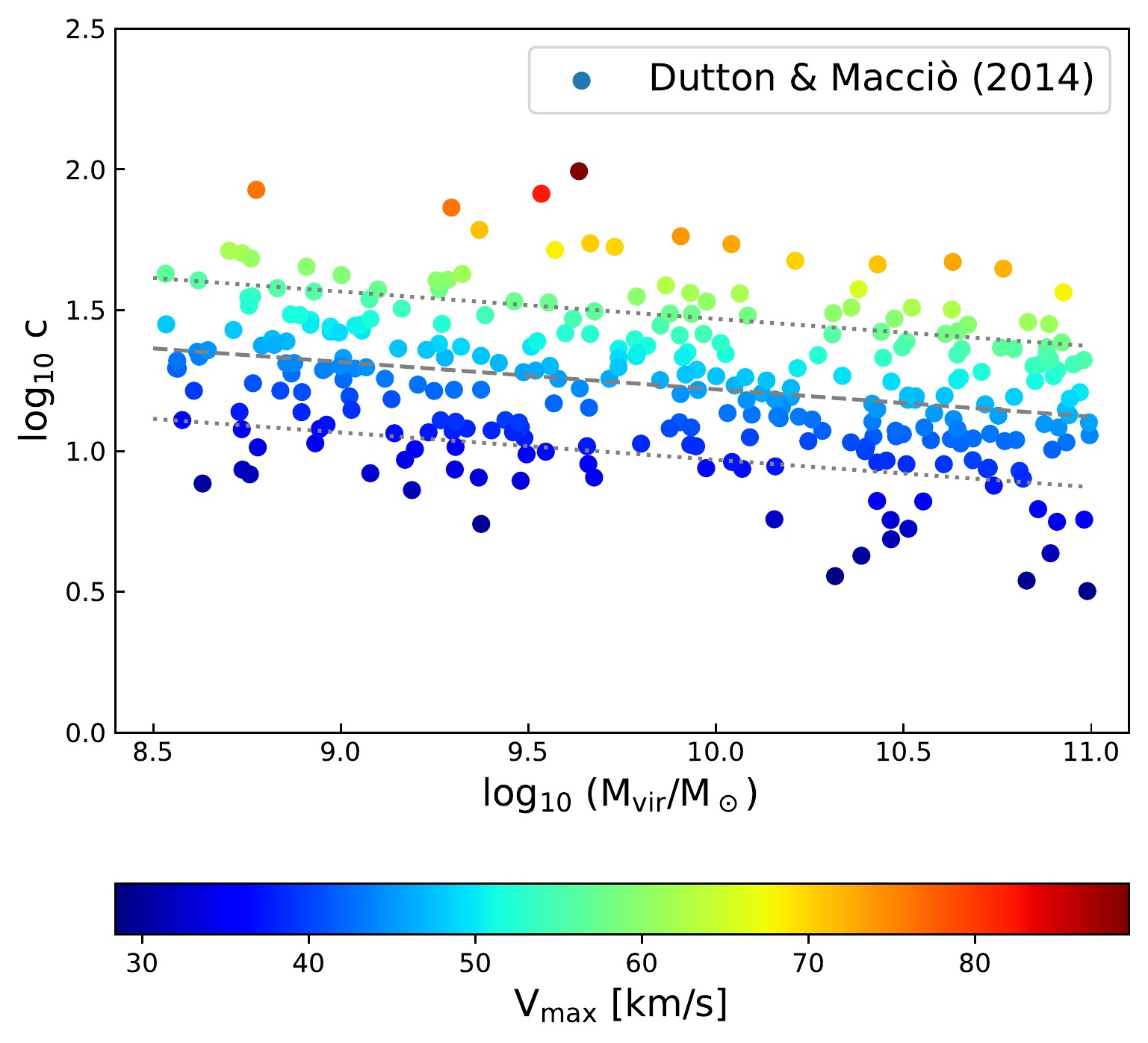}
		\caption{Distribution of NFW halos in virial mass - concentration plane used to construct circular velocity profiles in Figure \protect{\ref{fig:tbtf}}.  Halos are color coded by $V_{\rm max}$.
		}
		\label{fig:tsamp}
	\end{figure}

	The NFW circular velocity curves shown in Figure \ref{fig:tbtf} are generated for 300 mock halos which are uniformly sampled in virial mass between $8.5 \leq {\rm log} M_{\rm vir} \leq 11$ and follow the redshift zero halo mass concentration relation from \cite{Dutton14} with scatter of $\sigma_{{\rm ln} c} = 0.25$.  These halos are shown in Figure \ref{fig:tsamp}, color coded by their $V_{\rm max}$.  
	
\begin{figure}
		\centering
		\includegraphics[width=\hsize]{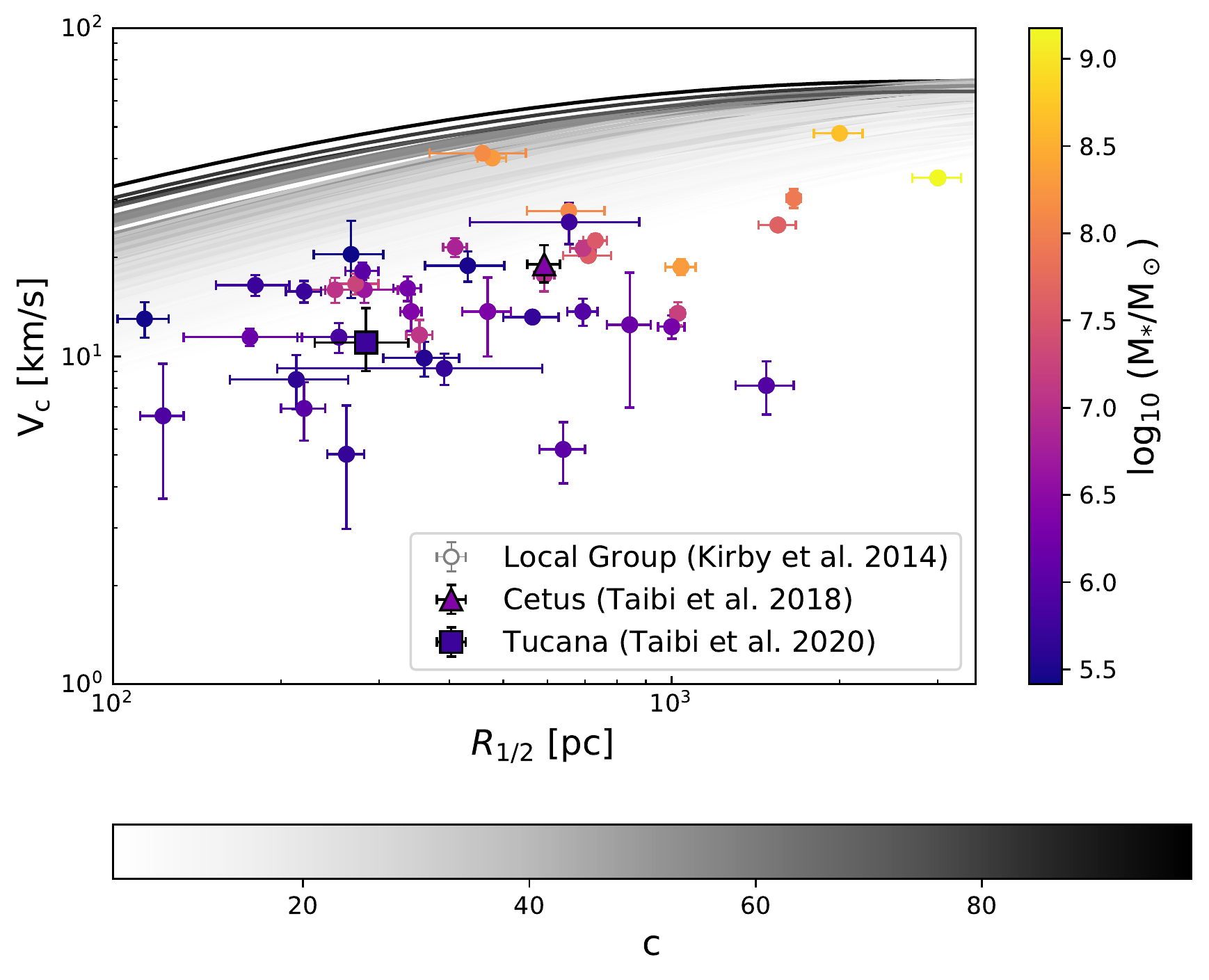}
		\includegraphics[width=\hsize]{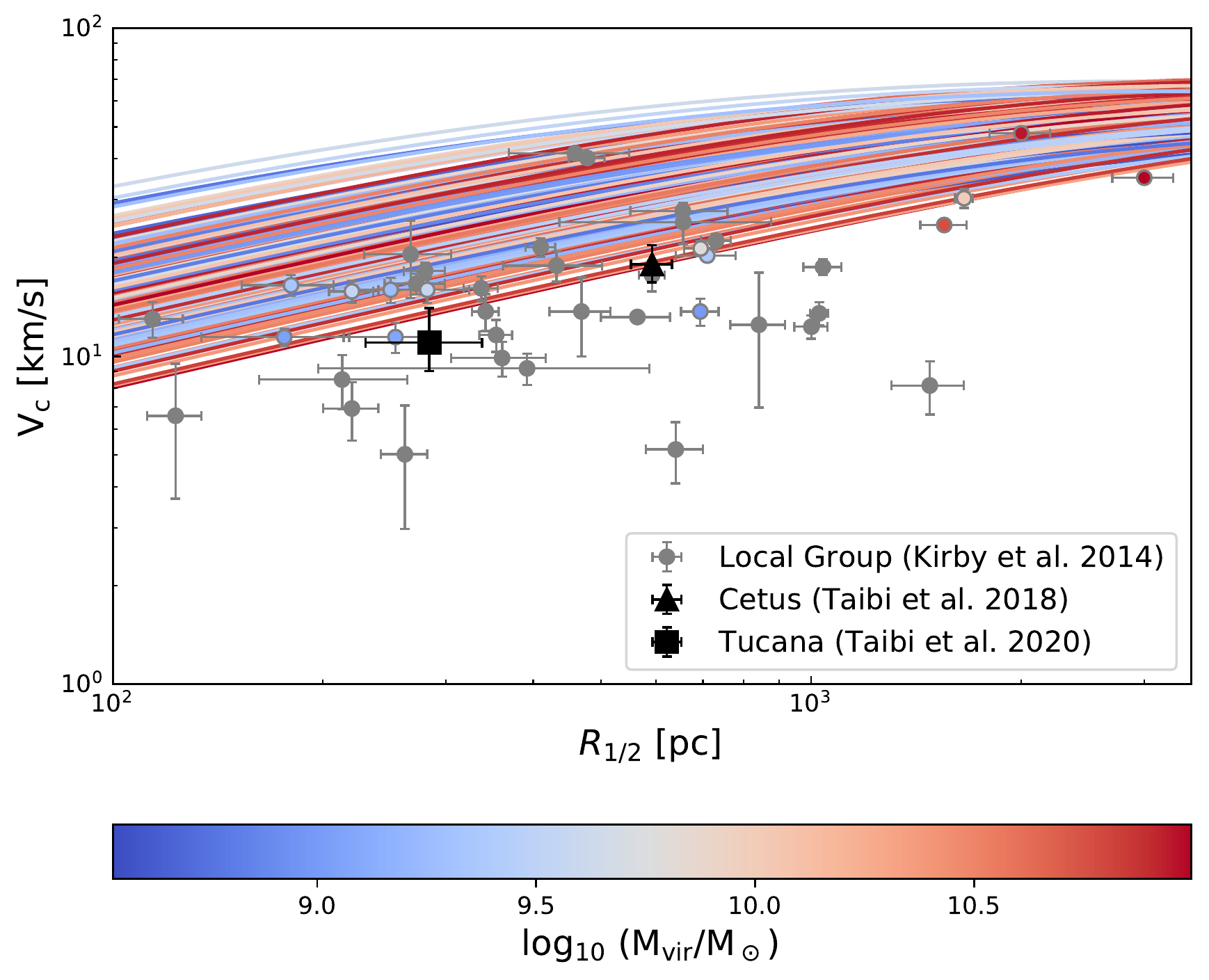}
		\caption{Reproduction of Figure \protect{\ref{fig:tbtf}}, but with halo circular velocity profiles color coded by concentration (\textit{top}) and virial mass (\textit{bottom}).  
		Cetus and Tucana are shown as a triangle and a square, respectively. 
		In the top panel, the Local Group dwarf galaxies are color coded accordingly to their stellar mass; a probable correlation is visible between the circular velocity and the stellar mass.
		In the bottom panel, the color coded dots that follow the same color schema as the halo virial mass profiles are Local Group dwarf galaxies for which virial masses were obtained from dynamical modeling \citep{Read19b,Leaman2012}.
		}
		\label{fig:tbtf2}
	\end{figure}

	In Figure \ref{fig:tbtf2} for completeness we reproduce plots where the mock halos are color coded by concentration, or virial mass.  
	In  the  top  panel,  the  Local  Group  dwarf  galaxies  are color coded accordingly to their stellar mass; a probable correlation is visible between the circular velocity and the stellar mass. In the bottom panel, observed dwarf galaxies with virial masses estimated from dynamical modeling \citep{Read19b,Leaman2012} are also color coded, illustrating the tension between the predicted and observed density profiles.
	
\end{appendix}	
	
\end{document}